\documentclass[a4paper,fleqn,usenatbib]{mnras}

\usepackage{newtxtext,newtxmath}
\usepackage[T1]{fontenc}
\usepackage{ae,aecompl}

\defcitealias{Citro2017}{\scshape C17}
\usepackage{graphicx}	% Including figure files
\usepackage{amsmath}	% Advanced maths commands
\usepackage{amssymb}	% Extra maths symbols

\usepackage{pdflscape}

\usepackage{xcolor}
\usepackage[normalem]{ulem}

\newcommand{\OH}{log([\ion{O}{iii}]/H$\alpha$) \,}
\newcommand{\NO}{log([\ion{N}{ii}]/[\ion{O}{ii}]) \,}

\title[Catching galaxies in the act of quenching star formation]{Galaxies in the act of quenching star formation}

\author[S. Quai et al.]{
Salvatore Quai,$^{1,2}$\thanks{E-mail: salvatore.quai@unibo.it}
Lucia Pozzetti,$^{2}$
Annalisa Citro,$^{1,2}$
Michele Moresco$^{1,2}$ 
\newauthor and Andrea Cimatti$^{1,3}$
\\
$^{1}$Dipartimento di Fisica e Astronomia, Universit\`a di Bologna, Via Gobetti 93/2, I-40129, Bologna, Italy\\
$^{2}$INAF - Osservatorio Astronomico di Bologna, Via Gobetti 93/3, I-40129, Bologna, Italy\\
$^{3}$INAF - Osservatorio Astrofisico di Arcetri, Largo E. Fermi 5, I-50125, Firenze, Italy\\}

\date{Accepted XXX. Received YYY; in original form ZZZ}

\pubyear{2017}

\begin{document}
\label{firstpage}
\pagerange{\pageref{firstpage}--\pageref{lastpage}}
\maketitle

\begin{abstract}
Detecting galaxies when their star-formation is being quenched is crucial to understand the mechanisms driving their evolution. We identify for the first time a sample of quenching galaxies selected just after the interruption of their star formation by exploiting the [\ion{O}{iii}]~$\lambda5007$/H$\alpha$ ratio and searching for galaxies with undetected [\ion{O}{III}]. 
Using a sample of $\sim174000$ star-forming galaxies extracted from the SDSS-DR8 at $0.04~\le~\text{z}~<~0.21$,we identify the $\sim300$ quenching galaxy best candidates with low [\ion{O}{iii}]/H$\alpha$, out of $\sim26~000$  galaxies without [\ion{O}{iii}] emission. They have masses between $\rm10^{9.7}$ and $\rm10^{10.8}\, M{_{\odot}}$,consistently with the corresponding growth of the quiescent population at these redshifts. Their main properties (i.e. star-formation rate, colours and metallicities) are comparable to those of the star-forming population, coherently with the hypothesis of recent quenching, but preferably reside in higher-density environments.Most candidates have morphologies similar to star-forming galaxies, suggesting that no morphological transformation has occurred yet. From a survival analysis we find a low fraction of candidates ($\sim$ 0.58\% of the star-forming population), leading to a short quenching timescale of $t_Q\sim$ 50~Myr and an \emph{e}-folding time for the quenching history of $\tau_Q\sim$ 90~Myr, and their upper limits of $t_Q< 0.76$ Gyr and $\tau_Q<$1.5 Gyr, assuming as quenching galaxies 50\% of objects without [\ion{O}{iii}]~($\sim7.5\%$).Our results are compatible with a 'rapid' quenching scenario of satellites galaxies due to the final phase of strangulation or ram-pressure stripping. This approach represents a robust alternative to methods used so far to select quenched galaxies (e.g. colours, specific star-formation rate, or post-starburst spectra).
\end{abstract}

\begin{keywords}
galaxies: evolution
galaxies: abundances
galaxies: general
ISM: HII regions
ISM: lines and bands
\end{keywords}

%%%%%%%%%%%%%%%%%%%%%%%%%%%%%%%%%%%%%%%%%%%%%%%%%%

%%%%%%%%%%%%%%%%% BODY OF PAPER %%%%%%%%%%%%%%%%%%

%%%%%%%%%%%%%%%%%%%%%%%%%%%%%%%%%%%%%%%%%%%%%%%%%%
%%%%%%%%%%%%%%%%%%%%%%%%%%%%%%%%%%%%%%%%%%%%%%%%%%
%%%%%%%%%%%%%%%%%%%%%%%%%%%%%%%%%%%%%%%%%%%%%%%%%%

%%%%%%%%%%%%%%%%%%%%%%%%%%%%
% INTRODUCTION
%%%%%%%%%%%%%%%%%%%%%%%%%%%%

\section{INTRODUCTION}
\label{sec:Intro}
Since the pioneering work of Hubble \citep{Hubble1926}, galaxies have been divided 
into two broad populations: blue star-forming spirals (late-type galaxies) and 
red ellipticals and lenticulars (early-type galaxies) with weak or absent star 
formation. The advent of massive surveys, such as the Sloan Digital Sky Survey 
\citep[SDSS,][]{York2000, Strauss2002}, provided very large samples of all galaxy
types and allowed to study their general properties with unprecedented statistics. 

At low redshifts (z $\sim 0.1$), galaxies show a bimodal distribution of their 
colours \citep{Strateva2001,Blanton2003, Hogg2003,Balogh2004,Baldry2004} 
and structural properties \citep{Kauffmann2003,Bell2012}. In a colour-magnitude (CMD) diagram or in a colour-mass diagram, early-type and bulge-dominated galaxies 
occupy a tight 'red sequence'. Instead, late-type, disk-dominated systems are 
spread in the so-called 'blue cloud' region. At higher redshifts, this bimodality 
has been clearly observed up to at least z $\sim 2$ \citep[e.g.][]{Willmer2006,Cucciati2006,
Cirasuolo2007, Cassata2008, Kriek2008,Williams2009,Brammer2009, Muzzin2013}.
The increase of the number density and the stellar mass growth of the red
population from $z\sim 1-2$ to the present 
\citep[e.g.][]{Bell2004,Blanton2006,Bundy2006,Faber2007,Mortlock2011,Ilbert2013,Moustakas2013} suggests that a fraction of blue galaxies migrates from the blue cloud to the red 
sequence, together with a transformation of their morphologies and the suppression 
of the star formation (quenching) \citep[e.g.][]{Pozzetti2010,Peng2010}. 
An interesting possibility is that both galaxy bimodality and the growth of the red 
population with cosmic time are due to a migration of the disk-dominated galaxies 
from the blue cloud to the red sequence when they experience the interruption (quenching) of the star 
formation while, at the same time, there is a continued assembly of massive 
(near L$^\ast$), red spheroidal galaxies through dry merging along the red sequence 
\citep{Faber2007,Ilbert2013}. It is also thought that these transitional scenarios depend on 
the environment where galaxies are located \citep[e.g.][]{Goto2003,Balogh2004,Peng2010}. 

Interestingly, the CMD region between the blue and the red populations is 
underpopulated \citep[e.g.][]{Balogh2004}. In particular, the distribution of 
optical colours, at fixed magnitude, can be fitted by the sum of two separate 
Gaussian distributions \citep{Baldry2004}, without the need for an intermediate 
galaxy population. This suggests that the transition timescale from star-forming to passive galaxies must be rather short \citep[e.g.][]{
Martin2007, Mendez2011, Mendel2013, Salim2014}. By considering also ultraviolet data, it has been possible to better explore the CMD at $z \sim 0.1$ thanks to colours more sensitive 
to young stellar populations (lifetimes $< 100$ Myr) with respect to the standard optical CMD 
\citep{Wyder2007, Martin2007, Salim2007, Schiminovich2007}. This showed that 
there is an excess of galaxies in a wide region between the red sequence and the 
blue cloud that is not easily explained with a simple superposition of the two 
populations. This intermediate region has been named 'green valley' and it should be populated by galaxies just in the process of interrupting their star formation (quenching).

The physical origin of the star formation quenching is still unclear, and 
many mechanisms have been proposed \citep[see][]{SomervilleDave2015}. 
The most appealing options include (i) the radiative and mechanical processes due to 
AGN activity \citep[e.g.][]{Fabian2012}, (ii) the quenching due to the gravitational 
energy of cosmological gas accretion delivered to the inner-halo hot gas by cold 
flows via ram-pressure drag and local shocks 
\citep[{\em gravitational quenching},][]{Dekel2008}, the suppression of star formation when a disk becomes 
stable against fragmentation to bound clumps without requiring gas 
consumption, the removal or termination of gas supply \citep[{\em morphological quenching},][]{Martig2009}, and the processes due to the interaction between 
the galaxy gas with the intracluster medium in high density environments \citep[{\em environmental or satellite quenching},][]{Gunn1972,Larson1980, Moore1998, Balogh2000, Bekki2009, Peng2010, Peng2012}.
The quenching processes are also termed internal or environmental depending on whether they are originated  
within a galaxy or if they are triggered by the influence of the environment (e.g. the 
intracluster medium). These processes are not mutually exclusive, and they could
in principle take place together on different timescales. For instance, the 
environmental quenching (e.g. gas stripping) is expected to be dominant only 
in dense groups and clusters. The internal AGN feedback and gravitational heating 
quenching are thought to be limited to halos with masses higher than $10^{12}$ 
M$_{\odot}$, whereas morphological quenching can play an important role also
in less massive halos and in field galaxies.

Despite the importance and necessity of quenching, the actual identification of
galaxies where the suppression of the star formation is taking place remains very
challenging. 
Several approaches have been exploited so far to find galaxies in the quenching phase.
In particular, most studies focused on galaxies migrating from the blue cloud to the 
red sequence. For instance, green valley galaxies show varied morphologies 
\citep{Schawinski2010a}, with a predominance of bulge-dominated disk shapes
\citep{Salim2014}. Furthermore, there is a consensus in interpreting the 
decreasing of the specific Star Formation Rate (sSFR) 
with redder colours \citep[e.g.][]{Salim2007,Salim2009,Schawinski2014} as an 
indicator of recent quenching or rapid decrease of the star formation. 
However, \cite{Schawinski2014} argue that, despite the lower sSFR, the green 
valley is constituted by a superposition of two populations that share the same 
intermediate optical colours: the green tail of the blue late-type galaxies with 
no sign of rapid transition towards the red sequence (quenching timescale of several 
Gyr) and a small population of blue early-type galaxies which are quickly transiting 
across the green valley (timescale $\sim 1$ Gyr) as a result of major mergers of 
late-type galaxies. 
The connection between quenching, red sequence and mergers has been 
investigated by studying the so called post-starburst (E+A or K+A) galaxies. These 
galaxies have morphological disturbances associated with gas-rich mergers and are 
spectroscopically characterized by strong Balmer absorptions (i.e. dominated by A type stars), although 
they do not show emission lines due to ongoing star-formation \citep[e.g.][]{Quintero2004,Goto2005,
Poggianti2004,Poggianti2008}. 
The strong Balmer absorption of post-starburst galaxies suggest that their star 
formation terminated 0.5-1 Gyr ago. 
Other interesting cases are represented by early-type galaxies with recently quenched 
blue stellar populations, or where star formation is occurring at low level and 
likely going to terminate soon \citep[e.g.][]{Thomas2010,Kaviraj2010,McIntosh2014}. Some 
recent results suggest that quenching may also occur with longer timescales during 
the inside-out evolution of disks and the formation of massive bulges via secular 
evolution \citep[e.g.][]{Tacchella2015,Belfiore2017}, or through the so called strangulation 
process \citep{Peng2015}. In addition to normal galaxies, many studies have been 
focused on systems hosting AGN activity in order to understand whether the released 
radiative and/or mechanical energy is sufficient to suppress star formation 
\citep[e.g.][]{Fabian2012}. Several results indicate that AGN feedback can indeed play a 
key role in the rapid quenching of star formation \citep[e.g.][]{Smethurst2016,Baron2017} 
and references therein). In the case of massive galaxies, the indirect evidence of 
a past rapid quenching ($< 0.5$ Gyr) of the star formation is also provided by the super-solar 
[$\alpha$/Fe] and the star formation histories (SFHs) derived for the massive early-type galaxies \citep[e.g.][]{Thomas2010,McIntosh2014,Conroy2016,Citro2016}. 

\begin{figure*}
   \centering
  \includegraphics[width=0.75\linewidth]{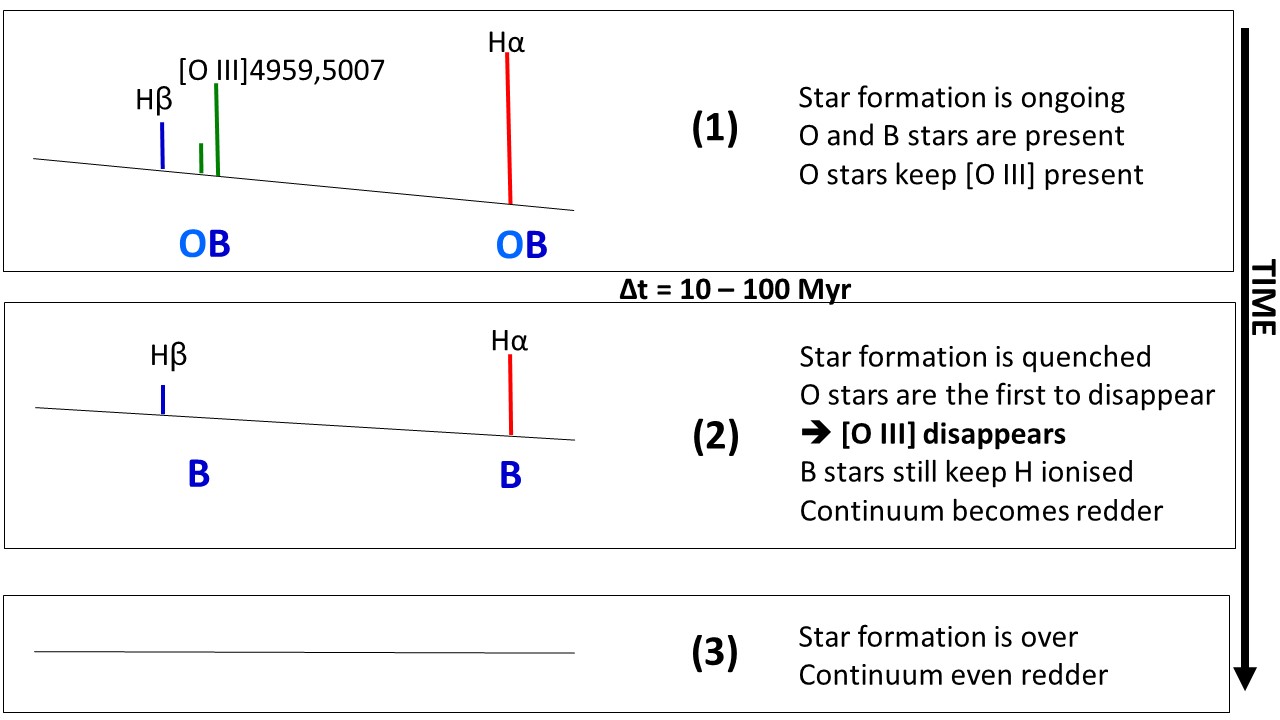}
   \caption{Scheme of the spectral evolution immediately after the quenching. By assuming a sharp quenching model by \protect\citetalias{Citro2017} the delay between phase (1) and (2) is $\sim$10 Myr and become $\sim$100 Myr by assuming a smoother and slower star formation decline (with a smoother exponential decline of the star-formation with $e$-folding time $\tau=200$ Myr). }
 \label{fig:Cimscheme}
\end{figure*}

To summarise, the results obtained so far identified galaxies some time
\emph{after} (e.g. post starbursts) or \emph{before} (e.g. AGN hosts) the 
quenching of the star formation. However, star-forming galaxies \emph{during} 
the quenching phase have not been securely identified. 
In this paper, we apply a new method aimed at selecting galaxies when their star formation is being quenched. Our search is
done at low redshift (0.04 $\leq$ z $<$ 0.21) within the SDSS main sample galaxy and is based on the use of higher (i.e. [\ion{O}{iii}]~$\lambda$5007, hereafter [\ion{O}{iii}]) to lower (mainly [\ion{O}{iii}]/H$\alpha$) ionisation emission line ratios. 
The modeling and details of the method have been presented by \citet[][hereafter C17]{Citro2017}. 
In principle, the selection of quenching galaxies is straightforward. 
After a few Myr from the interruption of star formation, the shortest-lived 
(i.e. most massive) O stars and their hard UV photons rapidly disappear, 
and this causes a fast decrease of the luminosity of high ionisation lines. 
However, emission lines with lower ionisation should be observable as long as 
late O and early B stars are still present before they abandon the main sequence (see \autoref{fig:Cimscheme}). Consequently, quenching galaxy candidates can be selected based on 
the high-to-low ionisation emission line ratios. 
However, the degeneracy between 
ionisation and maximum metallicity makes this approach less straightforward than it look, and additional
criteria should be used to identify the most reliable sample of quenching galaxies. In this
paper, we present the definition of the parent sample, the methods applied to mitigate the ionisation-metallicity degeneracy, the extraction of the most reliable quenching galaxy (QG hereafter) candidates, and the general properties of the selected sample.
We assume a flat $\Lambda$CDM cosmology with H$_0 = 70 \,\text{km s}^{-1}
\text{Mpc}^{-1}$, $\Omega_{\text{m}} = 0.3$ and $\Omega_\Lambda = 0.7$.

%%%%%%%%%%%%%%%%%%%%%%%%%%%%%%%%%%%%%%%%%%%%%%%%%%
%%%%%%%%%%%%%%%%%%%%%%%%%%%%%%%%%%%%%%%%%%%%%%%%%%
%%%%%%%%%%%%%%%%%%%%%%%%%%%%%%%%%%%%%%%%%%%%%%%%%%

%%%%%%%%%%%%%%%%%%%%%%%%%%%%
% THE SDSS SAMPLE GALAXY
%%%%%%%%%%%%%%%%%%%%%%%%%%%%
\section{THE SAMPLE SELECTION}

\subsection{The parent sample}
\label{subsec:sampsel}
Our sample is selected from the Sloan Digital Sky Survey Data Release 8\footnote{The 
data were downloaded from the CAS database, which contains catalogs of SDSS objects 
(\url{https://skyserver.sdss.org/CasJobs/})} \citep[SDSS-DR8,][]{Aihara2011}, adopting the following criteria:
\begin{enumerate}[(i)]
	\item keyword 'class'  = 'GALAXY';
	\item redshift range 0.04 $\leq$ z < 0.21;
	\item keyword 'LEGACY\_TARGET1' = 64
\end{enumerate}
The first criterion is clearly used to avoid stars and quasars. The second one
is adopted to minimise the biases due to the fixed aperture of SDSS spectroscopy. 
We identify systems where star formation is being quenched globally 
across the entire size of the galaxy. In order to ensure that the properties of 
the galaxy fraction measured inside the fibre aperture are reasonably representative 
of the global values, \cite{Kewley2005} found that the fibre should cover at least the 20 per cent of the observed B$_{26}$ isophote light of the galaxy. For SDSS, 
this fraction corresponds roughly to a redshift cut z > 0.04. 
Furthermore, we set an upper limit of $z < 0.21$, beyond which the number of objects rapidly decreases and do not significantly contributes to the sample statistic.

The third criterion ensures a homogeneous selection focusing on the Main Galaxy Sample
\citep[see][for details]{Strauss2002}, therefore avoiding mixing galaxies with different selection criteria. 

With the application of these three criteria, our total sample is constituted by 513596 galaxies. This sample is called {\em parent sample} hereafter, and includes all galaxy types
(from passive systems to star-forming objects) as well as Type 2 AGNs.
In order to avoid the spectral contamination due to sky lines residuals, 
we exclude $\sim 62400$ objects for which the centroids of the main emission lines 
(i.e. [\ion{O}{ii}]~$\lambda 3727$ -~hereafter [\ion{O}{ii}], H$\beta$,[\ion{O}{iii}], H$\alpha$ and [\ion{N}{II}]~$\lambda$6584 -~hereafter [\ion{N}{ii}]) are overlapped with the strongest sky lines. 

The spectral line measurements and physical parameters of the selected galaxies 
are obtained from the database of the Max Planck Institute for Astrophysics and the 
John Hopkins University (MPA-JHU measurements\footnote{see 
\url{http://wwwmpa.mpa-garching.mpg.de/SDSS/}.}). In particular, we retrieve the following quantities:
\begin{itemize}
	\item \emph{Emission lines flux}. The fluxes are measured with the technique 
described in \citet{Tremonti2004}, which is based  on the subtraction of the 
\citep[][BC03]{Bruzual2003} best-fitting population model of the 
stellar continuum, followed by a simultaneous fit of the nebular emission lines with a Gaussian profile. 
	\item \emph{Uncertainty in emission lines fluxes}. We use the updated 
uncertainties provided by \cite{Juneau2014}, which are obtained comparing statistically the 
emission line measurements of the duplicate observations of the same galaxies.  
	\item \emph{Stellar mass}. The stellar masses are estimated through 
SED fitting to the SDSS \emph{ugriz} galaxy photometry, using a Bayesian approach 
to a BC03 model grid. The 
magnitudes are corrected for the contribution of the nebular emission lines assuming that these contributions to the broad-band magnitudes \emph{u,g,r,i,z} are
the same inside and outside the 3$^{\prime\prime}$ fibre of the SDSS spectrograph. 
The obtained estimates are referred to the region sampled by the fibre. 
To obtain the total stellar mass, the MPA-JHU group corrected the stellar masses 
with a factor obtained by the difference between fibre magnitudes and total magnitudes. 
For this work, we assume the \emph{total stellar masses} corresponding to the median of 
their Bayesian probability distribution function. 
	\item \emph{Rest-frame absolute magnitude}. The rest-frame absolute magnitudes 
are derived from the \emph{ugriz} broad-band photometry, and corrected for the AB magnitude system. 
	\item \emph{Nebular Oxygen abundance}. Nebular oxygen abundance are estimated 
using a Bayesian approach, adopting the \cite{Charlot2001} models as discussed 
in \citet{Tremonti2004} and \cite{Brinchmann2004}. 
The estimates of oxygen 
abundances are expressed in 12~+~log(O/H) values, and were derived only when the signal-to-noise ratio S/N in H$\alpha$, H$\beta$, [\ion{O}{iii}] and [\ion{N}{II}] is > 3. In this work, we consider the 12~+~log(O/H) value corresponding 
to the median of the Bayesian probability distribution function.
	\item \emph{EW(H}$\alpha$\emph{)}. We adopt the rest-frame 
equivalent widths estimated by the SDSS pipeline with a continuum corrected for emission lines.
	\item \emph{D$_\text{n}4000$}. We use the D$_\text{n}4000$ \citep{Balogh1999} 
corrected for emission lines contamination.
	\item \emph{Galaxy size and light concentration}. The size is represented by 
the radius enclosing the 50 percent of the $r$-band Petrosian flux (R50). The light 
concentration is defined as C=R90/R50, where R90 is the radius containing the 90 percent 
of the $r$-band Petrosian flux. 
	\end{itemize}
In addition to these quantities, we also collect information about the galaxy
\emph{environment} by cross-matching our sample with the catalog provided by \cite{Tempel2014}, that contains environmental information relative to SDSS-DR10\footnote{The catalogue is available 
at \url{http://cosmodb.to.ee}} and we found a match for about the $87\%$ of galaxies in our sample. 
In particular, we use the environmental density they provide for each galaxy ($\rho_\text{env}$, hereafter), which represents an estimate of the overdensity %of the group/cluster whose the galaxy belongs to, 
with respect to the mean galaxy density within a scale of 1 h$^{-1}$ Mpc centered on each galaxy.
Furthermore, we use their \emph{Richness} and \emph{Brightness Rank}, that are defined, 
respectively, as the number of members of the group/cluster the galaxy belongs to,
and the luminosity rank of the galaxy within the group/cluster.

Finally, we analyze the SDSS morphological probability distribution of the galaxies provided by \cite{Huertas-Company2011}\footnote{We downloaded the SDSS morphological probability distribution of the galaxies together with the \cite{Tempel2014} catalogue.}, which is built by associating a probability to each galaxy of belonging to one of four morphological classes (Scd, Sab, S0, E).  

\subsection{The H$\alpha$ emission subsample}

As anticipated in the Section~\ref{sec:Intro}, our aim is to identify galaxies in the critical 
phase when the star formation is being suppressed. In the case of an instantaneous quenching (see \autoref{fig:Cimscheme}), this translates in searching for star-forming galaxies 
where high-ionisation lines (e.g. [\ion{O}{iii}]) are suppressed due to the 
disappearance of the most massive O stars, whereas Balmer emission lines are still
present because their luminosity decrease more slowly due to 
photo-ionisation from lower mass (longer lived) OB stars.

For these reasons, we select a subsample of galaxies with H$\alpha$ emission considering the following criteria:
\begin{enumerate}[(i)]
	\item EW(H$\alpha$) and  EW(H$\beta$) $\leq$ 0, in order to select galaxies with Balmer emission lines.
	\item	S/N(H$\alpha$) $\geq$ 5. For the SDSS this corresponds to objects with
 H$\alpha$ fluxes above $\simeq 1.1 \times 10^{-16}$ erg s$^{-1}$ cm$^{-2}$.
        \item S/N(H$\beta$) $\geq$ 3, to be able to properly correct for dust extinction using
the H$\alpha$/H$\beta$ ratio. With this criterion, we exclude $\sim 6$\% of galaxies, 
and the corresponding limiting flux is $\simeq 3.2 \times 10^{-17}$ erg s$^{-1}$ cm$^{-2}$.	
\end{enumerate}

By construction, the H$\alpha$ sample includes galaxies with S/N(H$\alpha$) $\geq$ 5; however, the other emission lines can have a lower S/N. When a line flux has
S/N $ < 2$, we assign an upper limit to the flux defined as:  
\begin{equation}
	\text{F} \leq 2 \times \sigma_{\text{F}}; \label{eq:010616_1}.
\end{equation} 
In the case of [\ion{O}{iii}], which is the most important signature of quenching in our study,
this upper limit corresponds to F~$\lesssim 2.2 \times 10^{-17}$ erg~s$^{-1}$~cm$^{-2}$.

The full H$\alpha$ emission sample contains 244362 objects. Clearly, this sample includes a heterogeneous ensemble
of galaxies where emission lines are powered by different ionisation processes
(star-forming, type 2 AGNs, LINERs, etc.).
	
\subsection{The subsample of star-forming galaxies} \label{sec:SFsample}
\begin{figure*}
   \centering
   \includegraphics[width=0.8\linewidth]{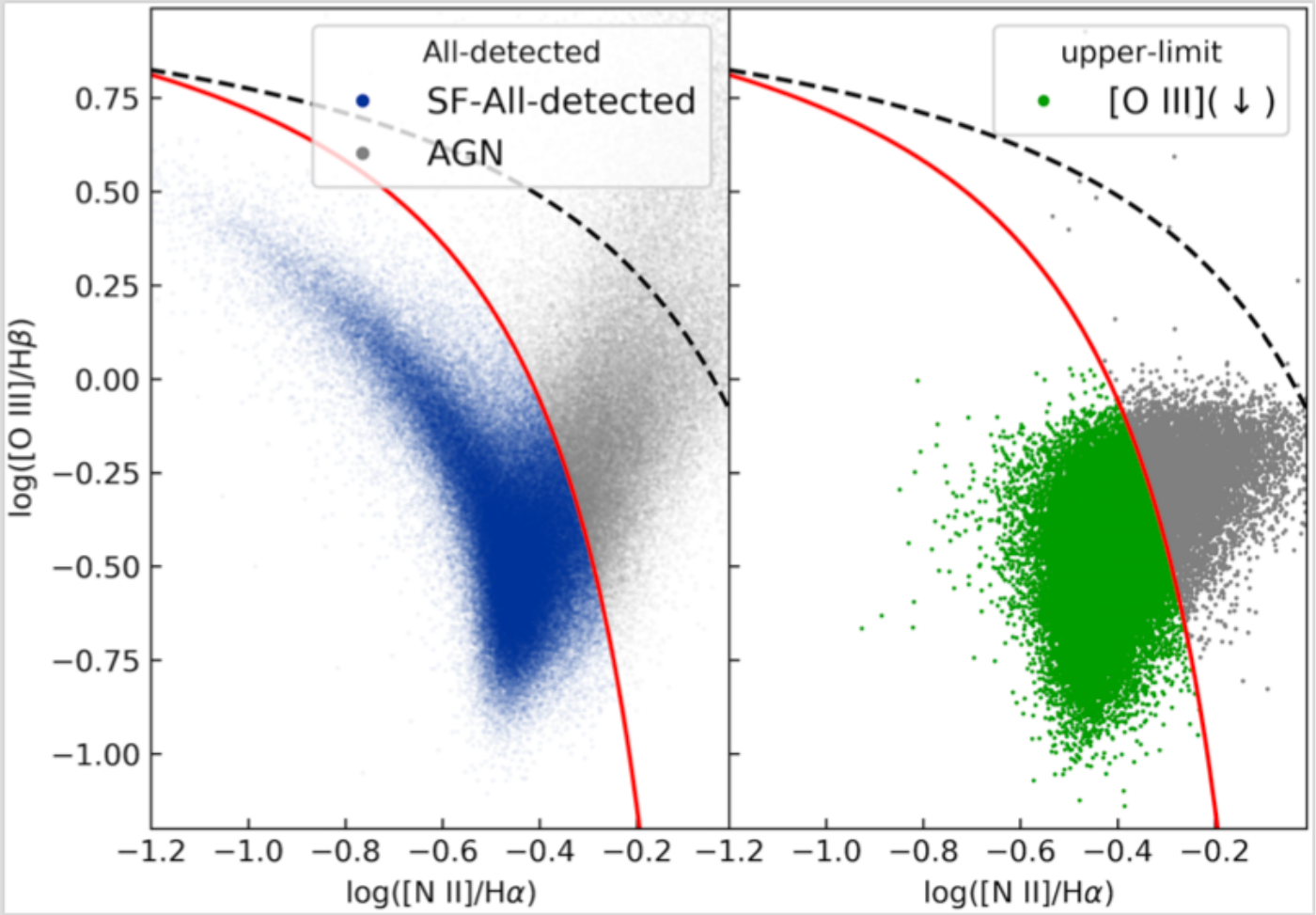}
   \caption{The BPT diagram of the H$\alpha$ subsample. 
     In the left panel we show the diagram for the galaxies whose all the BPT emission lines are detected (i.e. S/N > 5 for H$\alpha$, S/N > 3 for H$\beta$ and S/N > 2 for [\ion{O}{iii}] and [\ion{N}{ii}]).
    The red solid line is taken from \protect\cite{Kauffmann2003}, while the black dashed line was determined theoretically by \protect\cite{Kewley2001}.
    The blue dots represent the SF-Alldet galaxies, while the gray ones the AGNs.
     In the right panel we show the BPT diagram for galaxies with upper limit in [\ion{O}{iii}] line (i.e. galaxies with S/N([\ion{O}{iii}]) < 2). The green dots represent those galaxies that, despite their upper limit, satisfy the \protect\cite{Kauffmann2003} criterion for star-forming galaxies, while the gray ones represent objects for which we cannot be sure about their actual condition.
 }
         \label{fig:BPT_010616_1}
\end{figure*}

Since we are interested in purely star-forming systems, we cleaned the H$\alpha$
sample from contaminating galaxies. In order to separate the star-forming population 
from objects hosting AGN activity, we use the diagnostic diagram of \citet[][hereafter BPT]{Baldwin1981}.

\autoref{fig:BPT_010616_1} shows the BPT diagram of our sample. 
We remind that the emission lines involved in this diagram are close enough in wavelength that the correction for dust extinction is negligible.
We adopt the \cite{Kauffmann2003} criterion\footnote{
Their criterion is defined as $\text{log([\ion{O}{iii}]/H}\beta) < 0.61/\{\text{log([\ion{N}{II}]/H}\alpha) 
- 0.05\} + 1.3$.} to reject type 2 AGNs, LINERs and composite objects from the 
H$\alpha$ sample. 
For galaxies, where all lines are detected, the star-forming
population can be easily isolated, %, and we will refer to these galaxies as \emph{SF-Alldet}.
while 
for galaxies where [\ion{O}{iii}] is undetected (i.e. S/N[\ion{O}{iii}] < 2), 
we select only those galaxies whose upper limits of [\ion{O}{iii}] flux lie below
the \cite{Kauffmann2003} relation. With this approach, we exclude 62125 AGNs 
and LINERs, and obtain the final subsample of 174056 star-forming galaxies.  

Then, we divide this sample into two subsamples:
\begin{description}
	\item [\bf{SF-Alldet}] (148145 galaxies). These are star-forming (SF)
galaxies where all the main emission lines (H$\beta$, [\ion{O}{iii}], H$\alpha$ and [N~II]) 
are significantly detected.
We, therefore, reject all the objects with S/N([\ion{N}{ii}]) < 2 (i.e. 241 objects). 
	\item [\bf{SF-[\ion{O}{iii}] undetected}] (25911 galaxies).
    These galaxies differ from the previous ones for their [\ion{O}{iii}], that in this case is undetected (i.e. S/N < 2).
\end{description}

In order to compare the SF galaxies with the other galaxy types, we also select
a complementary subsample of galaxies without H$\alpha$ emission from the parent
sample (\autoref{subsec:sampsel}). 
Hereafter, the extracted sample is called {\bf no-H$\alpha$} 
subsample, and includes 201527 galaxies with $\text{S/N(H}\alpha) < 5$. 
\autoref{table:190616_1} summarizes the number of galaxies in the different 
subsamples, while \autoref{fig:281116_1} shows some of the main parameter distributions of the subsamples (i.e. redshift, masses and color excess).

\begin{table}
\caption{Numbers and median redshift of galaxies in the different subsamples.}          
\label{table:190616_1}      
\centering                          
\begin{tabular}{cccc}        
\hline                
Sample & Subsample & Number & median z \\    
\hline
SF-H$\alpha$ && 174056 & 0.08 \\
	& SF-Alldet & 148145 & 0.08  \\
	&SF-[\ion{O}{iii}]undet & 25911 & 0.12  \\
\hline
no-H$\alpha$ & & 201527 & 0.12  \\
\hline                 
Total  & & 375583 & 0.10 \\      
\hline                                   %inserts single line
\end{tabular}
\end{table}

\subsection{The correction for dust extinction}
\label{sec:DustC}
\begin{figure*}
   \centering
   \includegraphics[width=0.69\columnwidth]{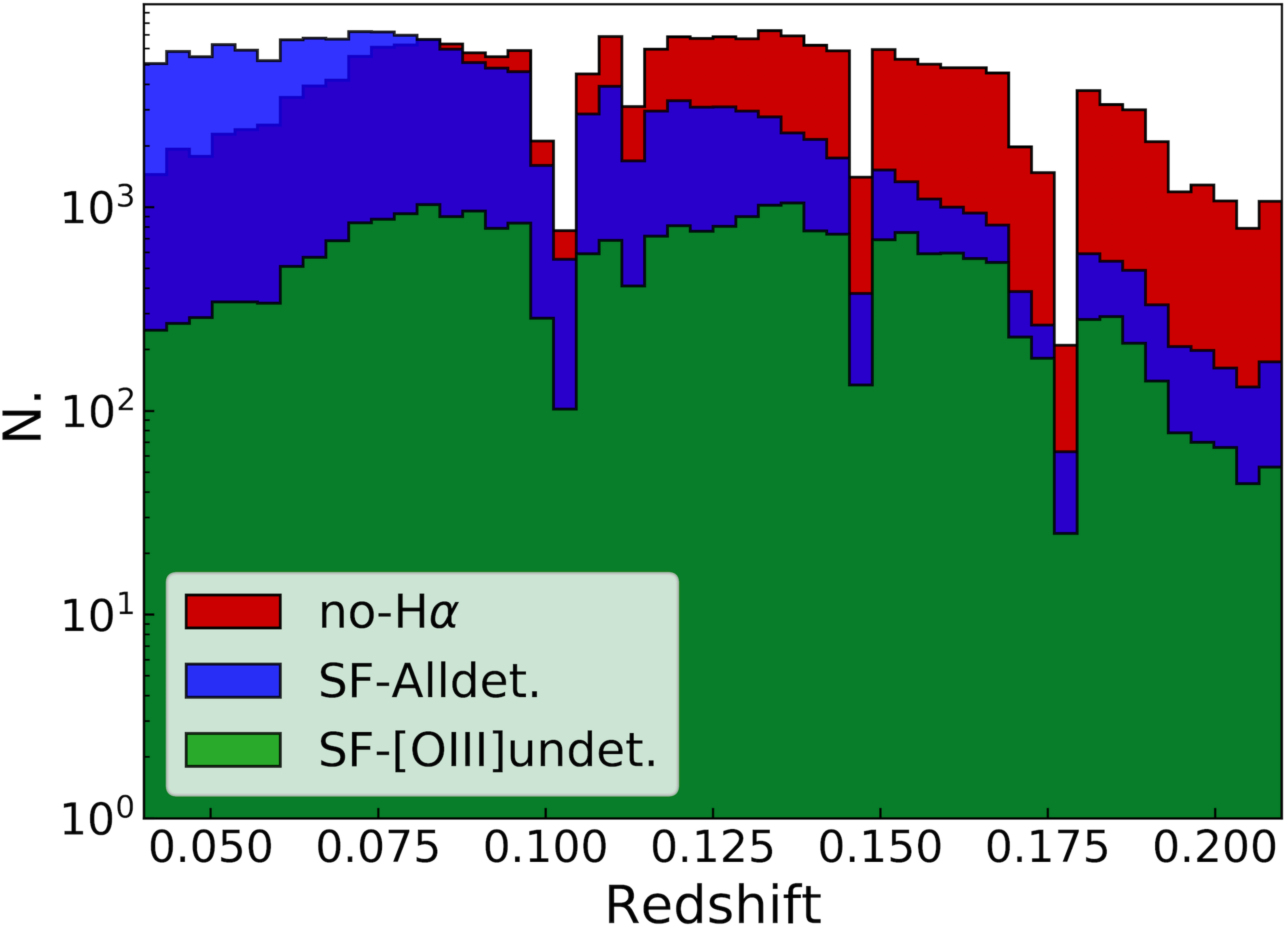} \hfill
  \includegraphics[width=0.69\columnwidth]{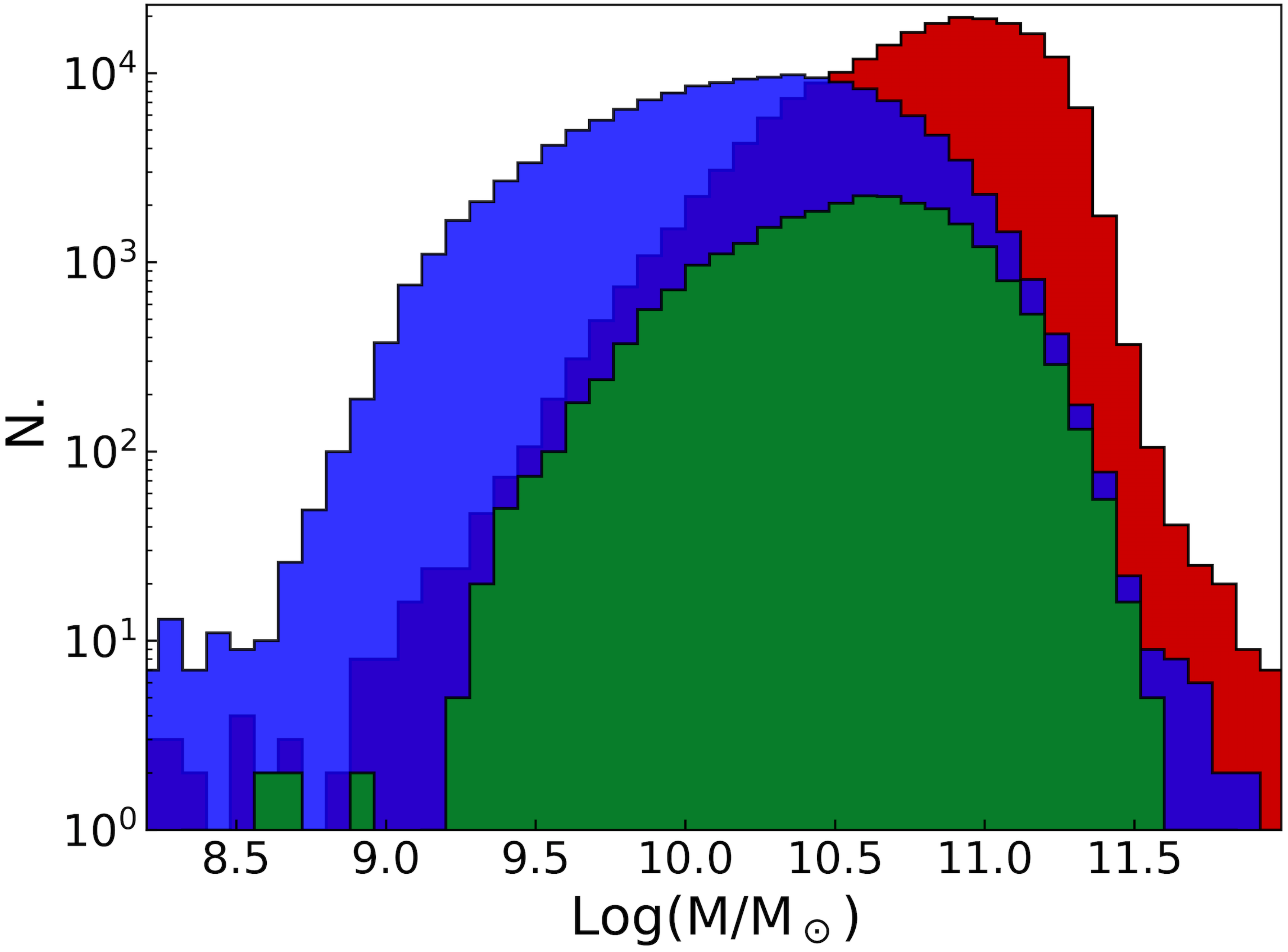} \hfill
\includegraphics[width=0.69\columnwidth]{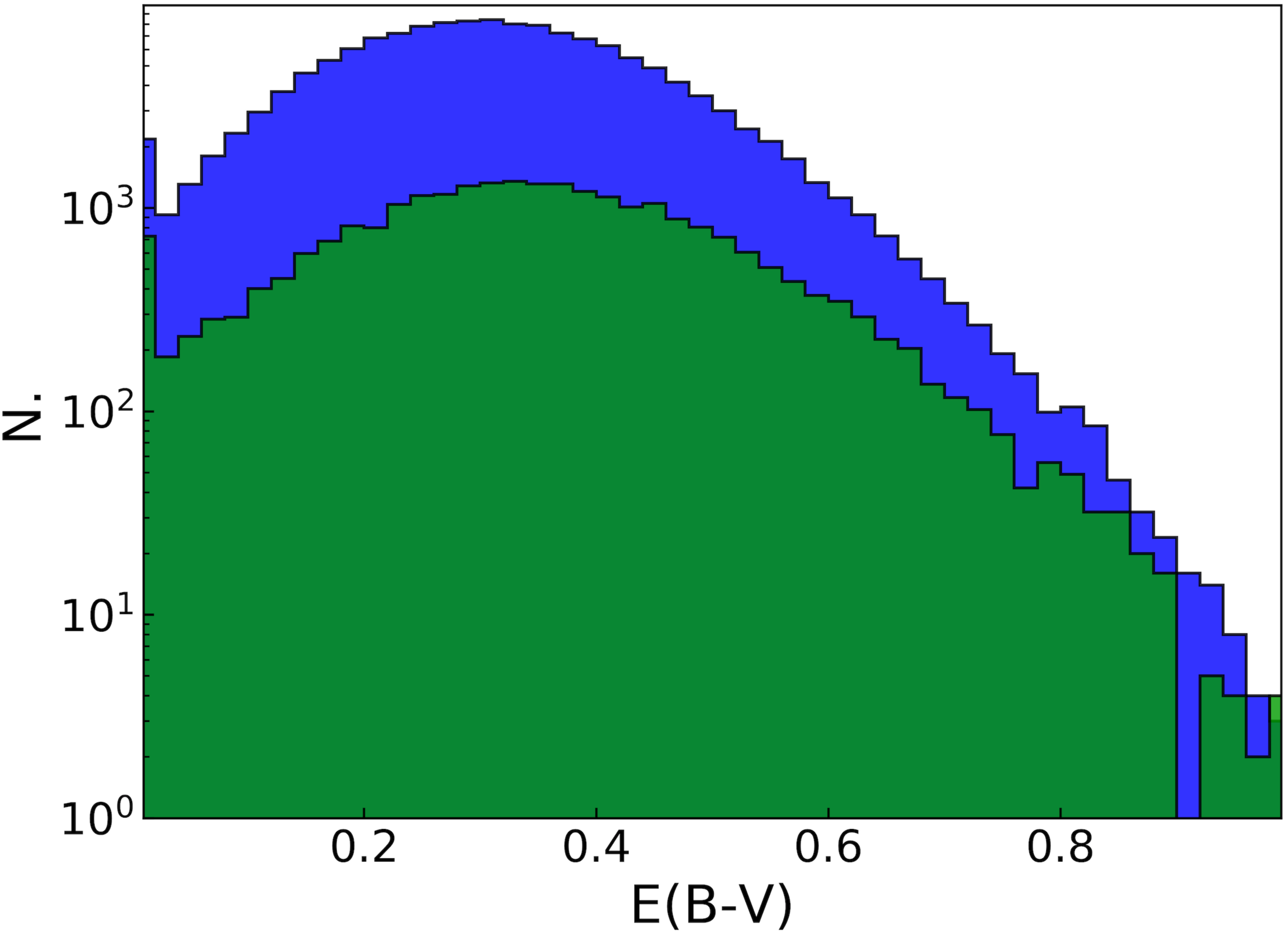} \hfill
      \caption{Main distributions of the three main subsamples: SF-Alldet in blue, SF-[\ion{O}{iii}]undet in green and no-H$\alpha$ in red: redshift (left panel), masses (central) and color excess E(B-V) (right).}
\label{fig:281116_1}
\end{figure*}
Since all the SF-Alldet and SF-[\ion{O}{iii}]undet galaxies have H$\beta$ with S/N$>$3, 
we correct their emission line fluxes for dust attenuation based on the 
H$\alpha$/H$\beta$ ratio, adopting the \cite{Calzetti2000} attenuation law.
The colour excess E(B-V) is derived assuming the Case B recombination and 
a Balmer decrement H$\alpha$/H$\beta = 2.86$ \citep[typical
of H II regions with electron temperature T$_\text{e} = 10^4$ K and electron 
density n$_\text{e} \sim 10^2-10^4 \,\text{cm}^{-3}$,][]{Osterbrock1989} 
For galaxies with H$\alpha$/H$\beta < $2.86  (2072 galaxies, $\sim$ 1\%), i.e. with a  negative colour excess, between $\sim -0.2 \leq \text{E(B-V)} <  0$, we 
decide to set E(B-V) = 0. The E(B-V) distribution is shown in \autoref{fig:281116_1}.

\subsection{The estimate of star formation rate}

After correcting the emission lines for dust extinction, we derive the star formation rates (SFRs) for the star-forming galaxies.  
The SFR is derived using the H$\alpha$ luminosity and adopting the 
\cite{Kennicutt1998} conversion factor for \cite{Kroupa2001} initial mass function 
(IMF): 

 \begin{equation}
 \text{SFR = L(H}\alpha)/10^{41.28} \;[\text{M}_\odot \text{yr}^{-1}]
 \end{equation}

In order to obtain the total SFRs, we correct the fibre SFRs for aperture effects.
Following \cite{Gilbank2010} and \cite{Hopkins2003} we apply an aperture correction $A$ 
based on the ratio of the \emph{u}-band Petrosian flux (which is a good approximation 
to the total flux) to the \emph{u}-band flux measured within the fibre:

\begin{equation}
A = \frac{\text{f}_\text{tot}(u)}{\text{f}_\text{fib}(u)} = 10^{-0.4(u_\text{fib} - u_\text{tot})}
\end{equation}

This method provides results which are in good agreement with the measurements of 
\cite{Brinchmann2004} obtained with a more complex approach \citep{Salim2007}.

%%%%%%%%%%%%%%%%%%%%%%%%%%%%%%%%%%%%%%%%%%%%%%%%%%
%%%%%%%%%%%%%%%%%%%%%%%%%%%%%%%%%%%%%%%%%%%%%%%%%%
%%%%%%%%%%%%%%%%%%%%%%%%%%%%%%%%%%%%%%%%%%%%%%%%%%

%%%%%%%%%%%%%%%%%%%%%%%%%%%%
% FINDING THE QUENCHING GALAXIES
%%%%%%%%%%%%%%%%%%%%%%%%%%%%

\section{FINDING THE QUENCHING GALAXIES}
\label{sec:catching}
\begin{figure*}
   \centering
  \includegraphics[width=0.75\linewidth]{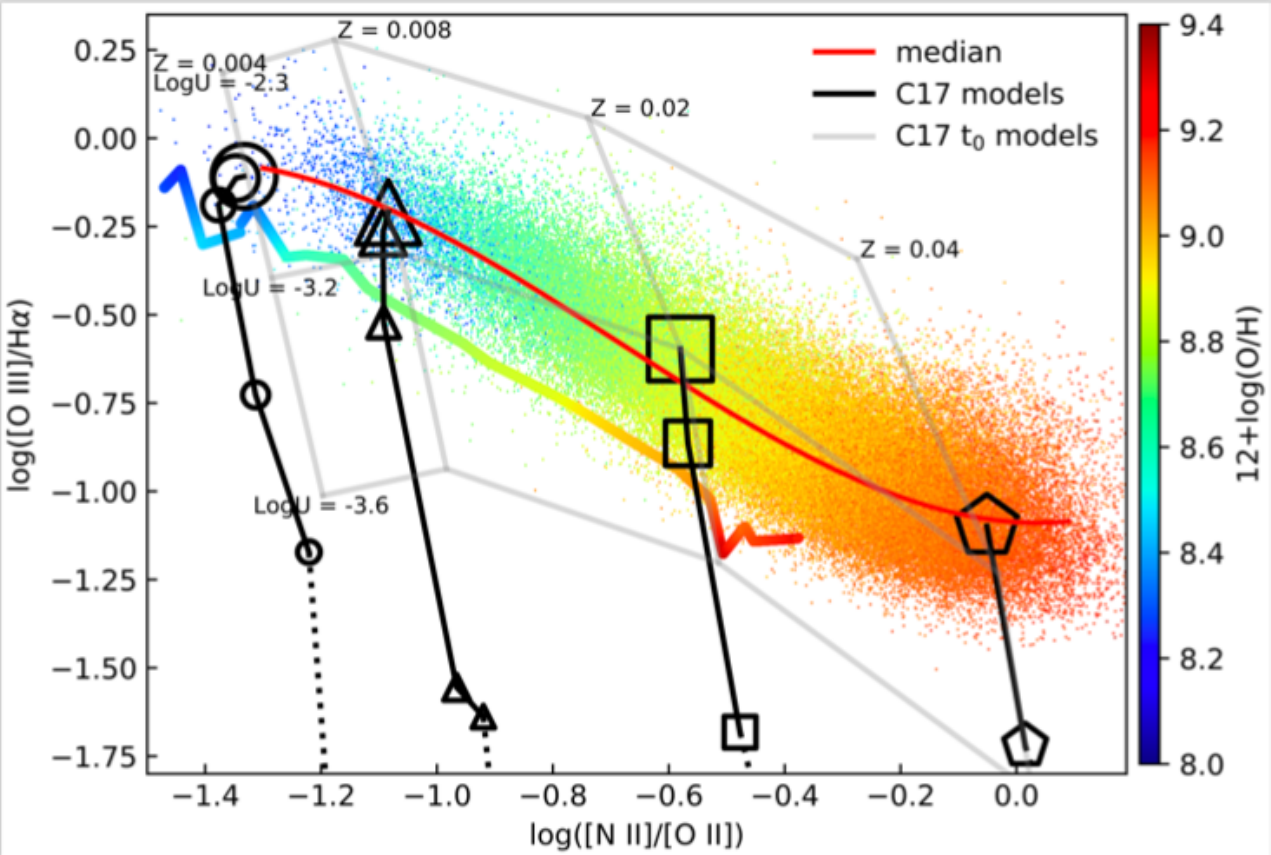}
   \caption{\OH vs. \NO diagram. The colour of the dots shows the metallicity estimate 12+log(O/H) by \protect\cite{Brinchmann2004} and the colorful line represents the dispersion at $3\sigma$ of the \NO for a given value of $12$+log(O/H). The red curve represents the median of the relation. Superimposed in gray is reported a grid of zero-ages \citetalias{Citro2017} models with different metallicities and ionisation parameters, while in black are shown the evolutive tracks by \citetalias{Citro2017} (circles for Z = 0.004, triangle for Z = 0.008, square for Z = Z$_\odot$ = 0.02, pentagon for Z = 0.04), where the size of the symbol is reported in time-step of 1 Myr with decreasing sizes.}
         \label{fig:121216_1_BIS}
\end{figure*}

\subsection{The general approach}
In this analysis, we decide to follow the approach discussed in \citetalias{Citro2017} to select galaxies in the phase when the quenching of their star formation takes place. In particular, \citetalias{Citro2017} showed how the ratio of high-ionisation (e.g. [\ion{O}{iii}] and [\ion{Ne}{iii}]~$\lambda 3869$ -~hereafter [\ion{Ne}{iii}]) to low-ionisation (e.g. Balmer lines) lines can be used to identify galaxies as close as possible to the time when the star formation starts to cease. 
Here, we explore in particular the  dust corrected [\ion{O}{iii}]/H$\alpha$ ratio (see \autoref{sec:DustC}) to select quenching galaxies. 
This ratio is highly sensitive to the ionisation parameter U and hence to the star formation evolutionary phase. 
In particular, higher values of U correspond to higher ionisation and star-formation levels \citepalias[for a more extensive discussion, see][]{Citro2017}. 
However, \OH is also dependent on the metallicity Z of the ionising stellar population, in the sense that low \OH values can be reproduced with both low U or high Z (i.e. ionisation-metallicity degeneracy, Z-U hereafter).
In order to find QG candidates, it is therefore necessary to mitigate this degeneracy. To address this issue, we devise two independent methods that are described in the following sections.

\subsection{Method A}
\label{sec:methA}
To mitigate the Z-U degeneracy, we firstly need to find an estimator for the metallicity independent of [\ion{O}{iii}].
Following \citetalias{Citro2017}, we exploit the [\ion{N}{II}]/[\ion{O}{ii}] ratio as metallicity indicator, as suggested e.g. by \cite{Nagao2006}.
\autoref{fig:121216_1_BIS} shows the dust corrected [\ion{O}{iii}]/H$\alpha$ vs. [\ion{N}{II}]/[\ion{O}{ii}] diagram as a function of the metallicity 12+log(O/H). In this analysis we discard 7712 objects with [\ion{O}{ii}] undetected (i.e. S/N [\ion{O}{ii}] < 2).

\autoref{fig:121216_1_BIS} clearly shows a very good correlation between [\ion{N}{II}]/[\ion{O}{ii}] and metallicity, with the advantage of having an almost orthogonal dependence between metallicity and log~U with respect to the BPT diagram, as confirmed also by the \citetalias{Citro2017} models shown in the figure. This allows to %\sout{more easily break} 
reduce the Z-U degeneracy, since, at fixed [\ion{N}{II}]/[\ion{O}{ii}], the spread of the distribution in [\ion{O}{iii}]/H$\alpha$  mainly reflects a difference in the ionisation status. 
For comparison, we also show the dispersion at $3\sigma$ of the \NO at a given 12+log(O/H).
In this diagram, therefore, at each [\ion{N}{II}]/[\ion{O}{ii}] (i.e. at fixed metallicity) galaxies with [\ion{O}{iii}]/H$\alpha$ lower than the $3\sigma$ dispersion due to metallicity can be considered as galaxies approaching the quenching, having an intrinsic lower ionisation parameter.

\begin{figure}
   \centering
   \includegraphics[width=\columnwidth]{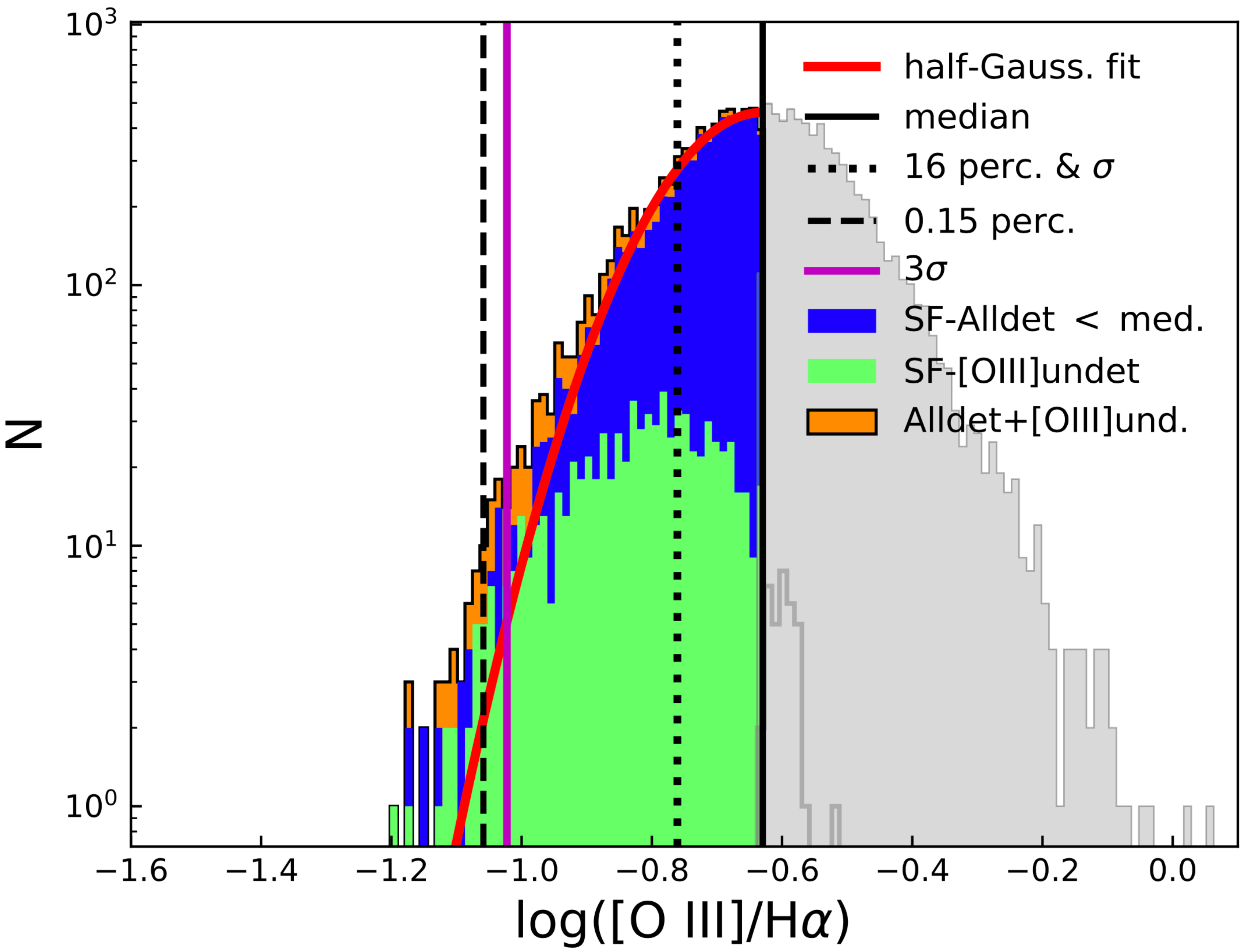}
   \includegraphics[width=\columnwidth]{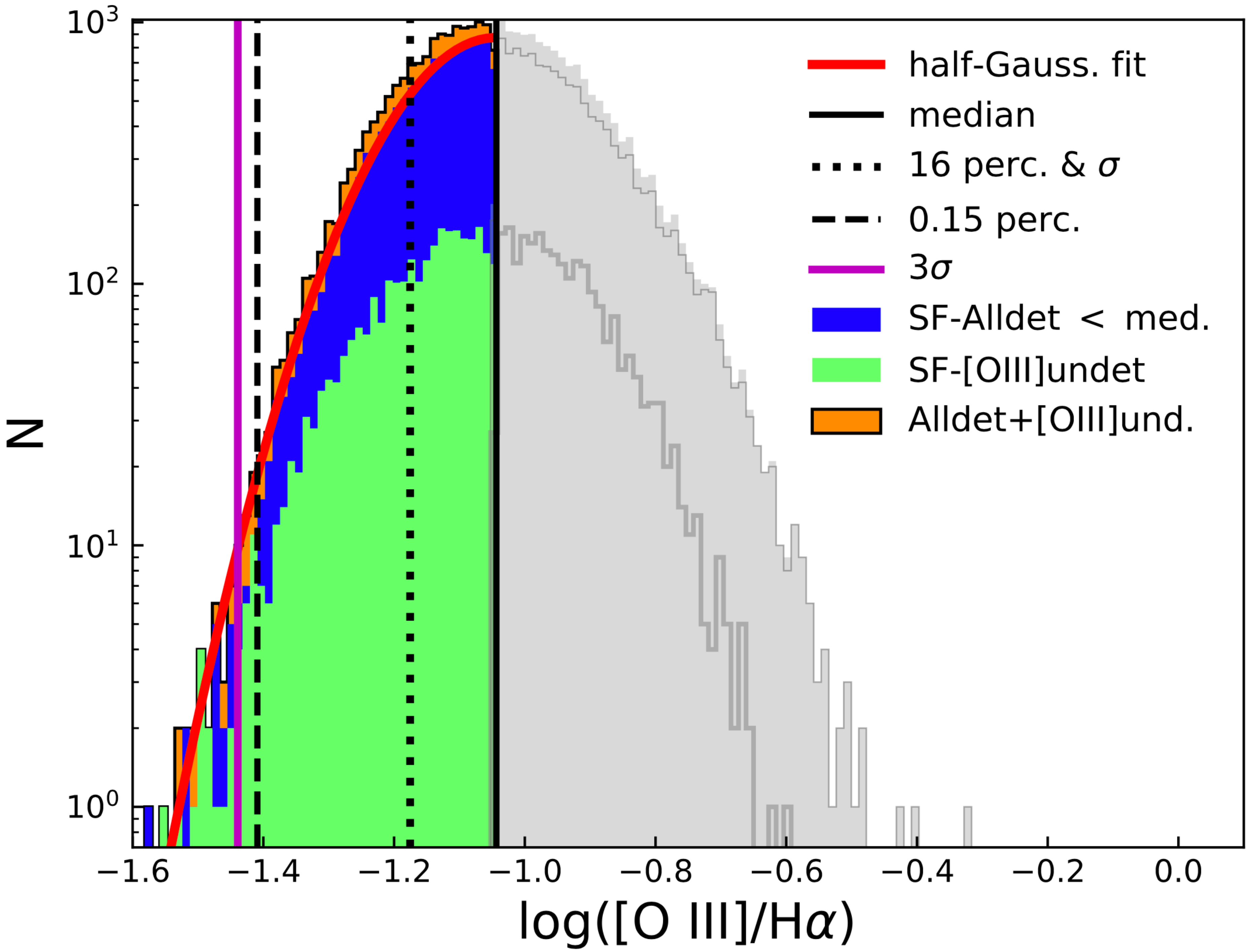}
   \caption{Distribution of [\ion{O}{iii}]/H$\alpha$ in bins of [\ion{N}{II}]/[\ion{O}{ii}]. For illustrative purposes, we show only two bins, namely $-0.7 \leq$~\NO $< -0.58$ and $-0.2 \leq$ \NO $< -0.08$.
The blue and gray histograms represent the distributions of the SF-Alldet galaxies respectively below and above the median, and the vertical lines are the median (solid line), the 16th percentile (dotted line) and  0.15th percentile (dashed line) of the distribution. The red solid line shows the Gaussian distribution obtained setting as mean and $\sigma$ the median and the 16th percentile of the distribution, respectively, and the purple vertical line represents the corresponding 3$\sigma$ of the half-Gaussian. An excess is identified when the median-3$\sigma$ of the Gaussian distribution is higher than the 0.15th percentile of the original distribution.
The green histogram represents the distribution of the SF-[\ion{O}{iii}]undet population and in orange  we show the contribution of these objects to the global distribution. 
}
         \label{fig:141216_1}
\end{figure}

The QG population should represent a population that separates from the SF sequence and starts transiting to the quenched phase. 
To isolate this extreme population, we analyse the SF-Alldet distribution of [\ion{O}{iii}]/H$\alpha$ in slices of [\ion{N}{II}]/[\ion{O}{ii}], searching for an excess of objects (using a Gaussian distribution as reference) with extremely low [\ion{O}{iii}]/H$\alpha$ values (i.e. lowest ionisation levels, see \autoref{fig:141216_1}). A similar approach was used to select starburst galaxies above the main sequence \citep{Rodighiero2011}. 
We focus in particular on the half part of the Gaussian distribution below the median, not considering the part above since it is dominated by ongoing SF, and could be biased by starburst systems and by a residual contamination of AGNs.

\begin{figure*}
   \centering
  \includegraphics[width=0.75\linewidth]{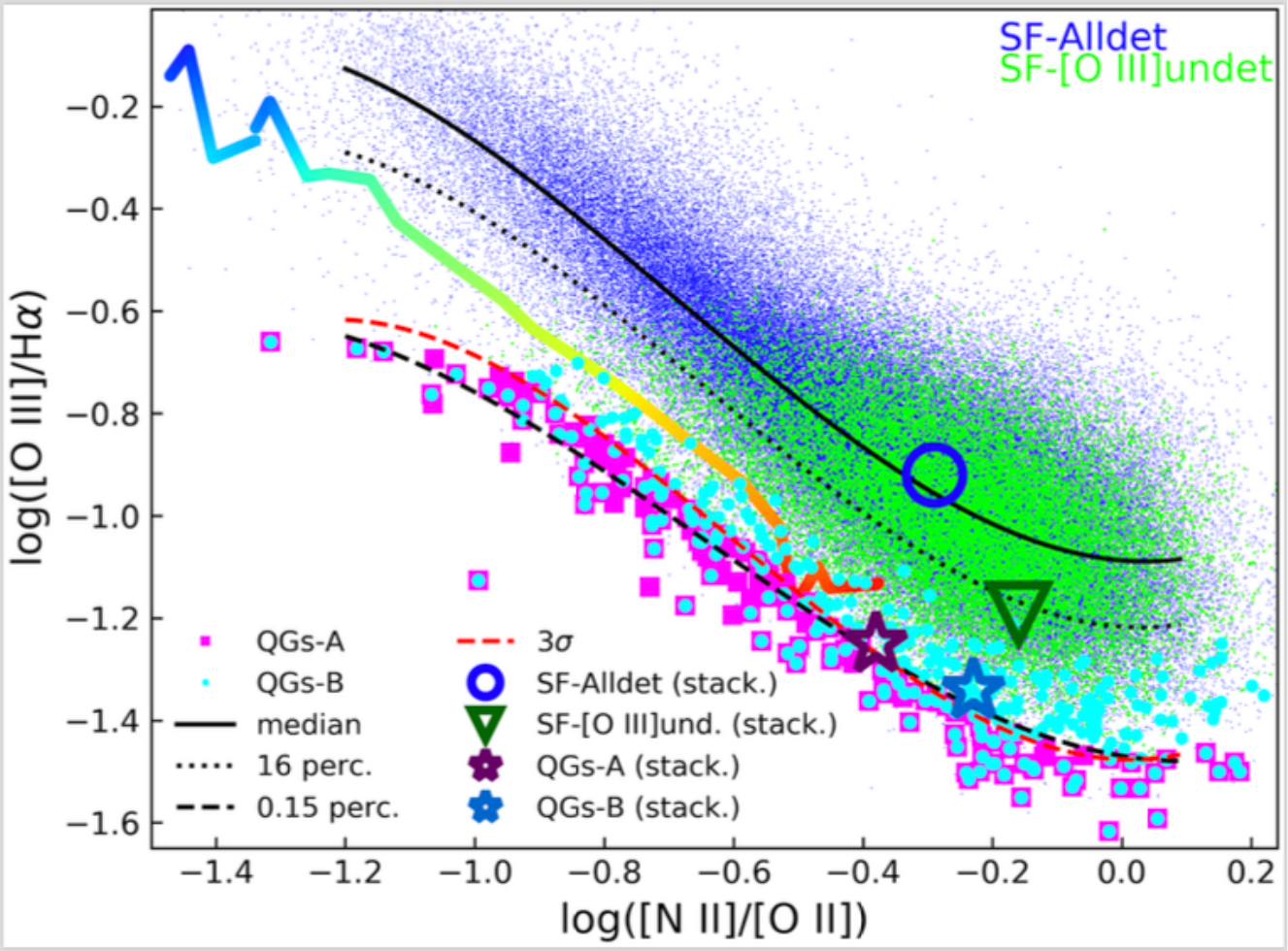}
   \caption{[\ion{O}{iii}]/H$\alpha$ vs. [\ion{N}{II}]/[\ion{O}{ii}] diagram. In blue are shown the SF-Alldet galaxies. The continue black curve represents the median of the relation, while the dotted and the dashed ones represent the 16th and the 0.15th percentiles, respectively. The dashed red curve represents the dispersion at $3\sigma$ of the relation. The superimposed colorful line represents the dispersion at $3\sigma$ of the \NO for a given value of $12$+log(O/H) (as in \protect\autoref{fig:121216_1_BIS}). In green, instead, are shown the SF-[\ion{O}{iii}]undet galaxies (i.e. upper limit in [\ion{O}{iii}]). Bigger dots represent the QG-A (magenta) and the QG-B (cyan). 
   The empty symbols represent the values estimated on the stacked spectra of SF-Alldet (blue circle), SF-[\ion{O}{iii}]undet (green triangle), QGs-A (purple star) and QGs-B (blue star) samples.}
         \label{fig:methodA_summ}
\end{figure*}

In detail, we proceed as follows:
\begin{itemize}
	\item We divide the distribution in bins of width $\Delta$[\ion{N}{II}]/[\ion{O}{ii}]$\approx$ 0.12 dex.
	\item In each bin of [\ion{N}{II}]/[\ion{O}{ii}], we estimate the median and the 16th percentile, and describe the distribution of SF-Alldet galaxies with an half-Gaussian whose mean and standard deviation ($\sigma$) are fixed to the median and the 16th percentile of the distribution.
	The typical value of $\sigma$ is $\sim 0.14$ dex. 
	\item We compare the 3$\sigma$ value of the Gaussian distribution with the 0.15th percentile, which corresponds to the 3$\sigma$ value in the case of a Gaussian distribution.
	\item When the median$-$3$\sigma$ value is higher than the 0.15th percentile, i.e. there is a positive detection of a deviation with respect to a Gaussian distribution, we identify our quenching galaxies as the excess beyond 3$\sigma$ %\sout{of galaxies} 
    with respect to the half-Gaussian (as shown in the upper panel of \autoref{fig:141216_1}).	
\end{itemize}
Following this approach, we find an excess of galaxies in the tail of the distribution in all bins with \NO $< -0.33$ (corresponding to 12+log(O/H)$\lesssim 9$). 
Above this threshold, instead, the limiting flux of our sample approaches \OH $\approx -1.5$, not allowing to detect candidates beyond the 3$\sigma$ value, as also shown in the bottom panel of \autoref{fig:141216_1}. 

To provide a less discrete description of the data, we generalise our method deriving the running median, the 16th percentile (representing also the $\sigma$ of the half-Gaussian), the 0.15th percentile and the median$-$3$\sigma$ for our SF-Alldet sample in the [\ion{O}{iii}]/H$\alpha$ vs. [\ion{N}{II}]/[\ion{O}{ii}] plane. We fit these relations with a third-order polynomial\footnote{Defined  $x =$ \NO) and $y = $ \OH, the polynomials are: $y(\text{median}) = -1.09 - 0.11x +1.39x^2 +0.67x^3$, $y(\text{16 perc.}) = -1.22 - 0.09x+1.48x^2 +0.76x^3$ and $y(\text{0.15 perc}) = -1.47 -0.21x +1.04x^2 + 0.53x^3$.}, and, to be more conservative, we define our QGs as the SF-[\ion{O}{iii}]undet galaxies lying below the median$-$3$\sigma$ polynomial of SF-Alldet population. 
This threshold is always below the dispersion in \NO at $3\sigma$ due to the metallicity (see \autoref{fig:methodA_summ}) and this suggests that the low \OH values are not related with the metallicity.

To further clean our sample, we discard also the 10 candidates that have [\ion{Ne}{iii}] detected, since it is a high-ionisation line and its presence is incompatible with the star-formation quenching  \citepalias[see][]{Citro2017}.

With this approach, we find 192 QG candidates (hereafter QG-A).
\autoref{fig:methodA_summ} shows the [\ion{O}{iii}]/H$\alpha$ vs. [\ion{N}{II}]/[\ion{O}{ii}] diagram, together with the selected QGs.

A possible issue with this method is that it is based on emission lines that are quite separated in wavelength, and therefore could be affected by  inappropriate correction for dust extinction. 
To test the impact of the extinction law on  [\ion{O}{iii}]/H$\alpha$ and [\ion{N}{II}]/[\ion{O}{ii}], we consider also the \cite{Seaton1979} extinction law instead of the \cite{Calzetti2000} one, finding a difference in the ratios at most of 0.1\%, and therefore not affecting strongly our selection. 

We also explore an alternative diagnostic diagram, considering the [\ion{O}{iii}]/H$\beta$ vs. [\ion{N}{II}]/[\ion{S}{ii}]\footnote{[\ion{S}{ii}] (i.e. [\ion{S}{ii}]$\lambda$6720) represents the sum of [\ion{S}{ii}]$\lambda$6717 and  [\ion{S}{ii}]$\lambda$6731 fluxes.}, discarding the redshift ranges in which the measurement of [\ion{S}{ii}] doublet could be biased by strong sky lines. 
The [\ion{N}{II}]/[\ion{S}{ii}] ratio has metallicity sensitivity similar to the one of  [\ion{N}{II}]/[\ion{O}{ii}], and [\ion{O}{iii}]/H$\beta$ is very similar to [\ion{O}{iii}]/H$\alpha$, with the drawback of  H$\beta$ being weaker than H$\alpha$.
This diagram has the advantage that the pairs of lines involved are close enough that it is possible to neglect the effect of dust extinction. 
Following the same procedure described above, we select 144 quenching candidates. 
However, since [\ion{S}{ii}] is intrinsically weaker than [\ion{O}{ii}], the S/N([\ion{S}{ii}]) distribution of these candidates peaks at S/N $\sim 2$ and, consequently, their identification is more uncertain.
We, therefore, decide not to consider them in the following. 

\subsection{Method B}
\label{sec:methB}

\begin{figure*}
   \centering
   \includegraphics[width=0.495\linewidth]{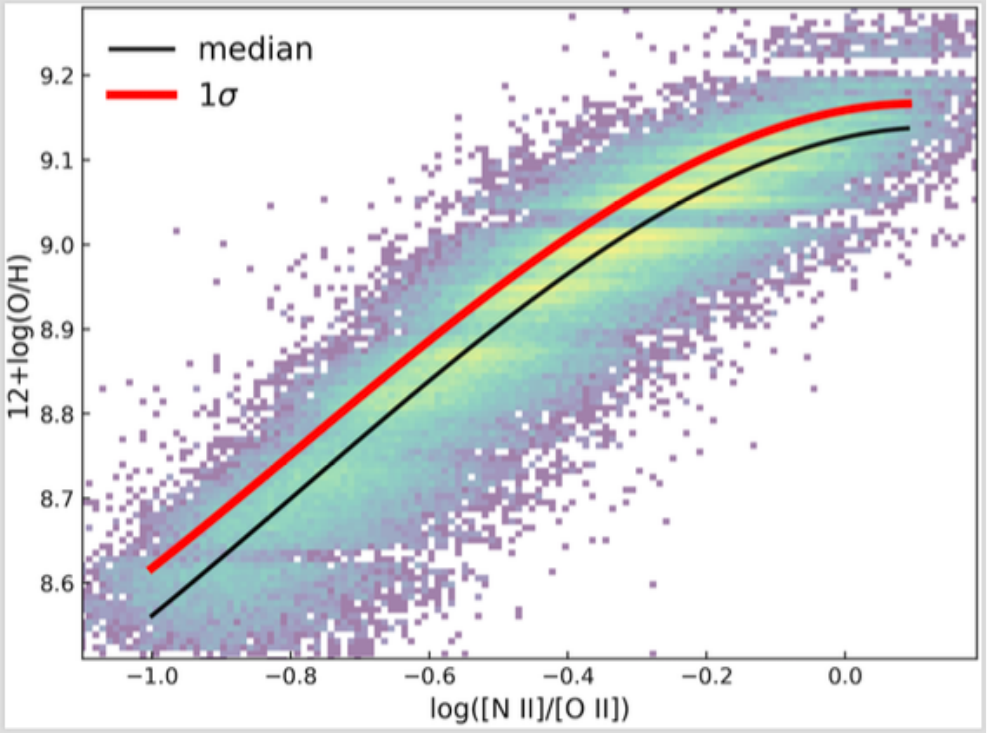}
  \includegraphics[width=0.495\linewidth]{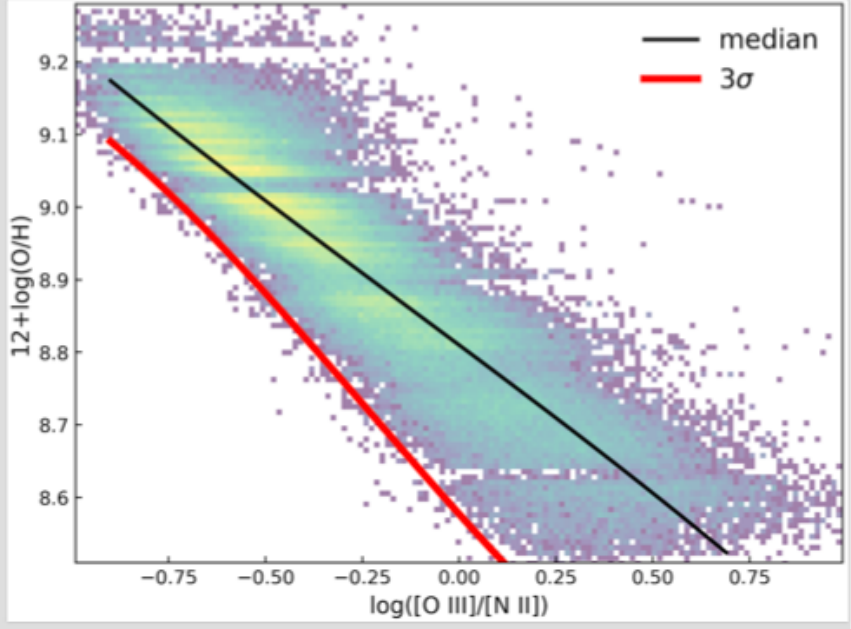}
   \caption{Left panel: 12+log(O/H) vs [\ion{N}{II}]/[\ion{O}{ii}] relation. The black curve represents the median of the relation, while the red curve represents the median +1$\sigma$, where $\sigma$ is the dispersion of the distribution. Right panel: 12+log(O/H) vs. [\ion{O}{iii}]/[\ion{N}{II}] relation. The black curve represents the median of the relation, while the red curve represents the median-3$\sigma$.}
         \label{fig:methodC_relat}
\end{figure*}

\begin{figure*}
   \centering
   \includegraphics[width=0.75\linewidth]{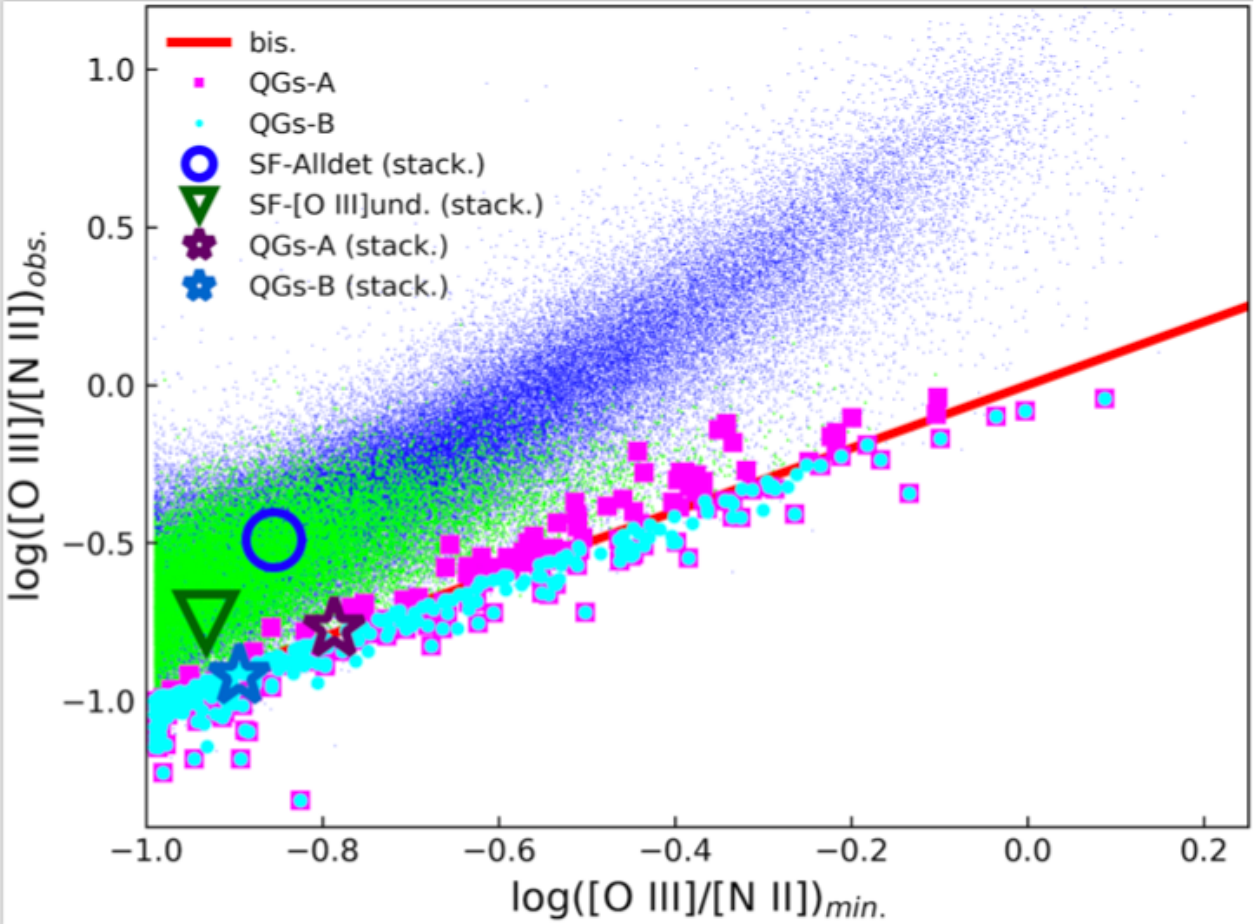}
   \caption{[\ion{O}{iii}]/[\ion{N}{II}]$_\text{obs.}$ vs. [\ion{O}{iii}]/[\ion{N}{II}]$_\text{min.}$ diagram. In blue are shown the SF-Alldet galaxies, while in green the SF-[\ion{O}{iii}]undet galaxies. Bigger dots represent the QG-A (magenta) and the QG-B (cyan). 
   The empty symbols represent the values estimated on the stacked spectra of SF-Alldet (blue circle), SF-[\ion{O}{iii}]undet (green triangle), QGs-A (purple star) and QGs-B (blue star) samples.}
         \label{fig:methodB_summ}
\end{figure*}

An alternative method to mitigate the Z-U degeneracy is to select galaxies for which [\ion{O}{iii}] is weaker than the minimum flux expected for the maximum metallicity. In this way, we can safely assume that the observed value of [\ion{O}{iii}]/H$\alpha$ is unlikely to be due to high-metallicity.
To investigate this possibility we proceed as follow:
\begin{itemize}
	\item  We derive a metallicity estimate (Z) for each galaxy in our sample. We exploit the 12+log(O/H) vs. [\ion{N}{II}]/[\ion{O}{ii}] relation suggested by \cite{Nagao2006} (see \autoref{fig:methodC_relat}), estimating the running median and the corresponding dispersion $\sigma_\text{Z}$ for the SF-Alldet sample. We therefore associate to each galaxy the median 12+log(O/H) corresponding to the observed [\ion{N}{II}]/[\ion{O}{ii}]\footnote{Note that we evaluate the expected metallicity also for galaxies with [\ion{O}{iii}] undetected, while \citet{Tremonti2004} derived metallicity only for galaxies with all emission lines detected %because of their weakness in some emission lines 
    (i.e. with the original S/N > 3).} as metallicity value.
	  \item We estimate the maximum metallicity (Z$_\text{max}$) of each galaxy, as:
		\begin{equation}
                      \text{Z}_\text{max} = \text{Z} + \sigma_\text{Z}, \label{eq:mB_1}
        		\end{equation}
where $\text{Z}$ and $\sigma_\text{Z} $ are the median and the dispersion associated to this relation (with $\sigma_\text{Z} \approx 0.13$ dex); since the dispersion would be higher than the typical metallicity errors, hence, $\text{Z}_\text{max}$ represents a statistical significant estimate for the maximum metallicity.

	\item In addition, we estimate the minimum expected [\ion{O}{iii}] flux for any given Z$_\text{max}$. To do that, we consider another relation between metallicity and emission lines that includes [\ion{O}{iii}].
    In particular, we adopt [\ion{O}{iii}]/[\ion{N}{II}] as a function of Z \citep{Nagao2006} (see \autoref{fig:methodC_relat}). This relation allows to estimate the minimum expected [\ion{O}{iii}]/[\ion{N}{II}] for any given Z$_\text{max}$: 
	\begin{equation}
	\text{[\ion{O}{iii}]/[\ion{N}{II}]}_\text{min.} = \text{[\ion{O}{iii}]/[\ion{N}{II}]} - 3\sigma_\text{[\ion{O}{iii}]/[\ion{N}{II}]},
	\end{equation}
where $\sigma_\text{[\ion{O}{iii}]/[\ion{N}{II}]}$ is the dispersion of the [\ion{O}{iii}]/[\ion{N}{II}]  vs. 12+log(O/H) relation (the typical dispersion is $\sigma_\text{[\ion{O}{iii}]/[\ion{N}{II}]} \approx 0.18$ dex).
	\item We identify as quenching candidates those galaxies with a [\ion{O}{iii}]/[\ion{N}{II}] lower than [\ion{O}{iii}]/[\ion{N}{II}]$_\text{min}$. For these galaxies, the low observed values of [\ion{O}{iii}]/[\ion{N}{II}] are unlikely due to their metallicity. Finally, as in method A, we discard $11$ galaxies with detected [\ion{Ne}{iii}], since it can be a sign of ongoing star formation.
\end{itemize}
In this way, we select 308 'method B' quenching candidates (hereafter QG-B), that are shown in
\autoref{fig:methodB_summ}. 

%%%%%%%%%%%%%%%%%%%%%%%%%%%%%%%%%%%%%%%%%%%%%%%%%%
%%%%%%%%%%%%%%%%%%%%%%%%%%%%%%%%%%%%%%%%%%%%%%%%%%
%%%%%%%%%%%%%%%%%%%%%%%%%%%%%%%%%%%%%%%%%%%%%%%%%%

%\section{Results}
\begin{figure}
   \centering   \includegraphics[width=\columnwidth]{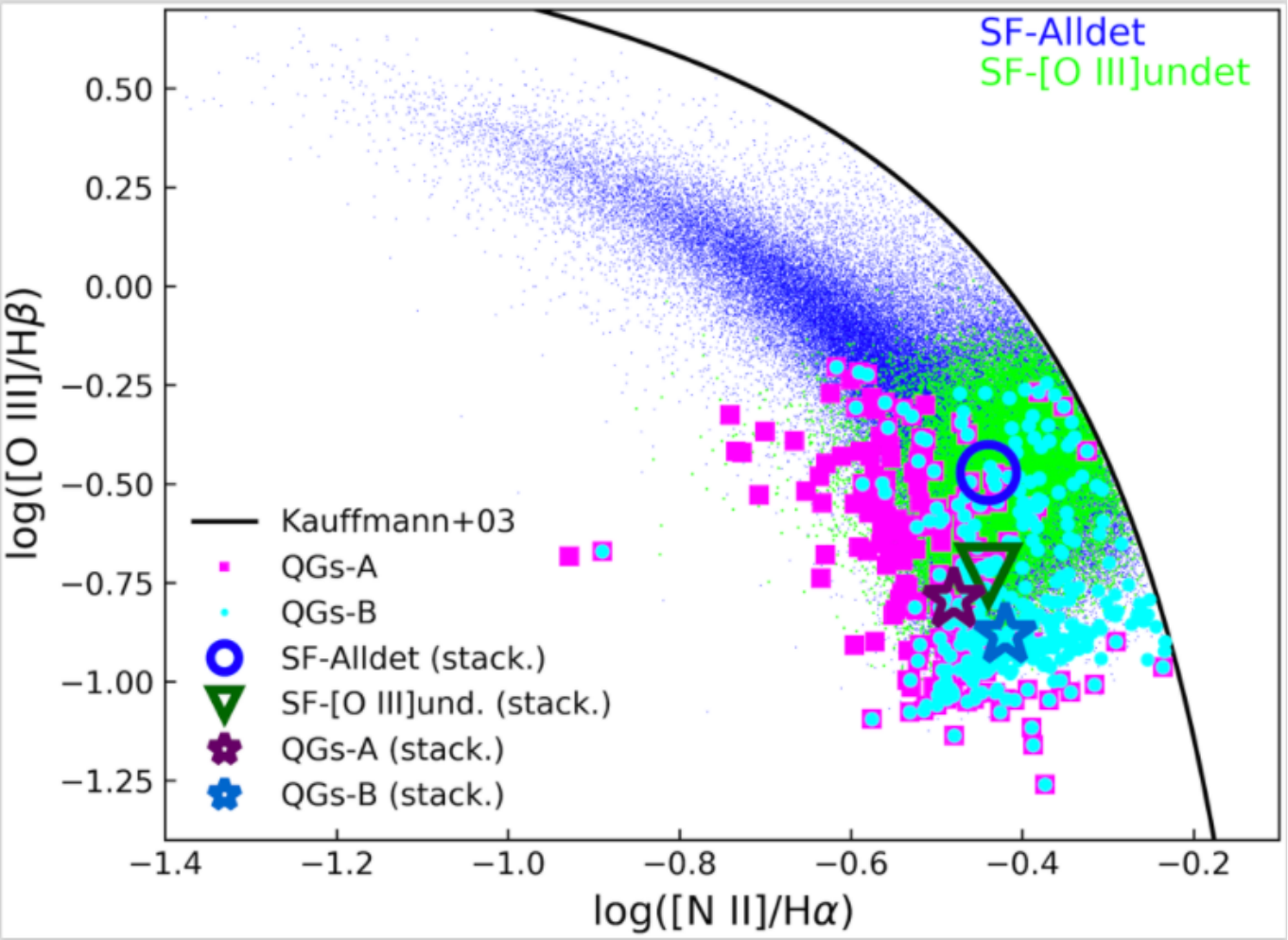}
     \caption{The BPT  diagram of our sample galaxies. The colour code is the same of \autoref{fig:methodB_summ}.}
         \label{fig:160117_2}
\end{figure}

%In Section~\ref{sec:catching} we have selected our quenching galaxy candidates by exploiting two methods. 

\subsection{Comparing the two methods}
In this section, we explore the differences and the complementarities between the two methods. % exploited, and the corresponding two sample selected.

We first notice that there are 120 QGs in common between the two methods.
In the [\ion{O}{iii}]/H$\alpha$ vs. [\ion{N}{II}]/[\ion{O}{ii}] diagram (i.e. the plane described in \autoref{sec:methA}, see \autoref{fig:methodA_summ}) there is a good agreement between Methods B and A: the bulk of QGs-B  are located below ($\sim$ 40\%) or close to the 3$\sigma$ curve that is the threshold criterion to select QGs-A. We also show the location of QGs-A candidates in the diagram used to select QGs-B (\autoref{fig:methodB_summ}). Also in this case there is a good agreement between the two samples, with QGs-A located below ($\sim$ 60\%) or just above the threshold criterion, having slightly higher upper limits for the observed [\ion{O}{iii}]/[\ion{N}{II}]. 
We could obtain a better agreement just slightly relaxing the thresholds adopted in the two methods. For example, if we adopt 2.5$\sigma$ as thresholds instead of 3 for both methods, we obtain that  about 60\% and 70\% of QGs-B are identified also as QGs-A and vice versa.
More conservative choices guarantee, however, to obtain more solid results and higher purity at the cost of a lower overlap.
Moreover, the residual discrepancy is due to an intrinsic difference between the two methods, that leads to select quenching candidates with different but complementary characteristics.
In \autoref{fig:160117_2} we show the BPT diagram with the candidates selected from the two methods. 
The bulk of QGs-A are distributed in the lower envelope of the BPT diagram at log([\ion{O}{iii}]/H$\beta) < 0$ and log([\ion{N}{II}]/H$\alpha) < -0.3$, while QGs-B are complementary located in a region at higher [\ion{N}{II}]/H$\alpha$ values (log([\ion{N}{II}]/H$\alpha) > -0.5$).
%{\color {green} This comparison suggest that, since QGs-B have higher values of [\ion{N}{II}]/H$\alpha$ and %lower values of 
%}[\ion{N}{ii}]/[\ion{O}{II}] {\color {green} 
%than QGs-A, method-B is able to identify, on average, higher metallicity QGs than method-A.}

%pre CITRO: This comparison, suggest that method-B, having QGs-B higher values of [\ion{N}{II}]/H$\alpha$ and lower values of [\ion{O}{iii}]/[\ion{N}{II}] compared to QGs-A, are 

%\begin{figure}
 %  \centering
 %  \includegraphics[width=\columnwidth]{immagini/CAND_BothMeth_O3Ha_vs_z.png} \hfill
  % \caption{ }
    %     \label{fig:200117_1}
%\end{figure}

\section {The properties of quenching galaxies}

In this section we analyse the %fundamentals 
properties of the QGs in order to identify or to constrain plausible quenching mechanism. 

In \autoref{tab:060217_1} we report the median, the 16th and 84th percentiles of the distribution of the main properties of our QGs, compared with those of the three control samples defined in Section~\ref{sec:SFsample}: SF-Alldet, SF-[\ion{O}{iii}]undet and no-H$\alpha$. 
We first notice that the median and the range in redshift of QGs candidates are similar to that of star forming galaxies (SF-Alldet).  
On the contrary, SF-[\ion{O}{iii}]undet and no-H$\alpha$ sample cover different redshifts ranges.

\subsection{Spectral properties}
\label{sec:spec_prop}
\begin{figure*}
   \centering
   \includegraphics[trim=0 0 0 0, clip, width=\linewidth]{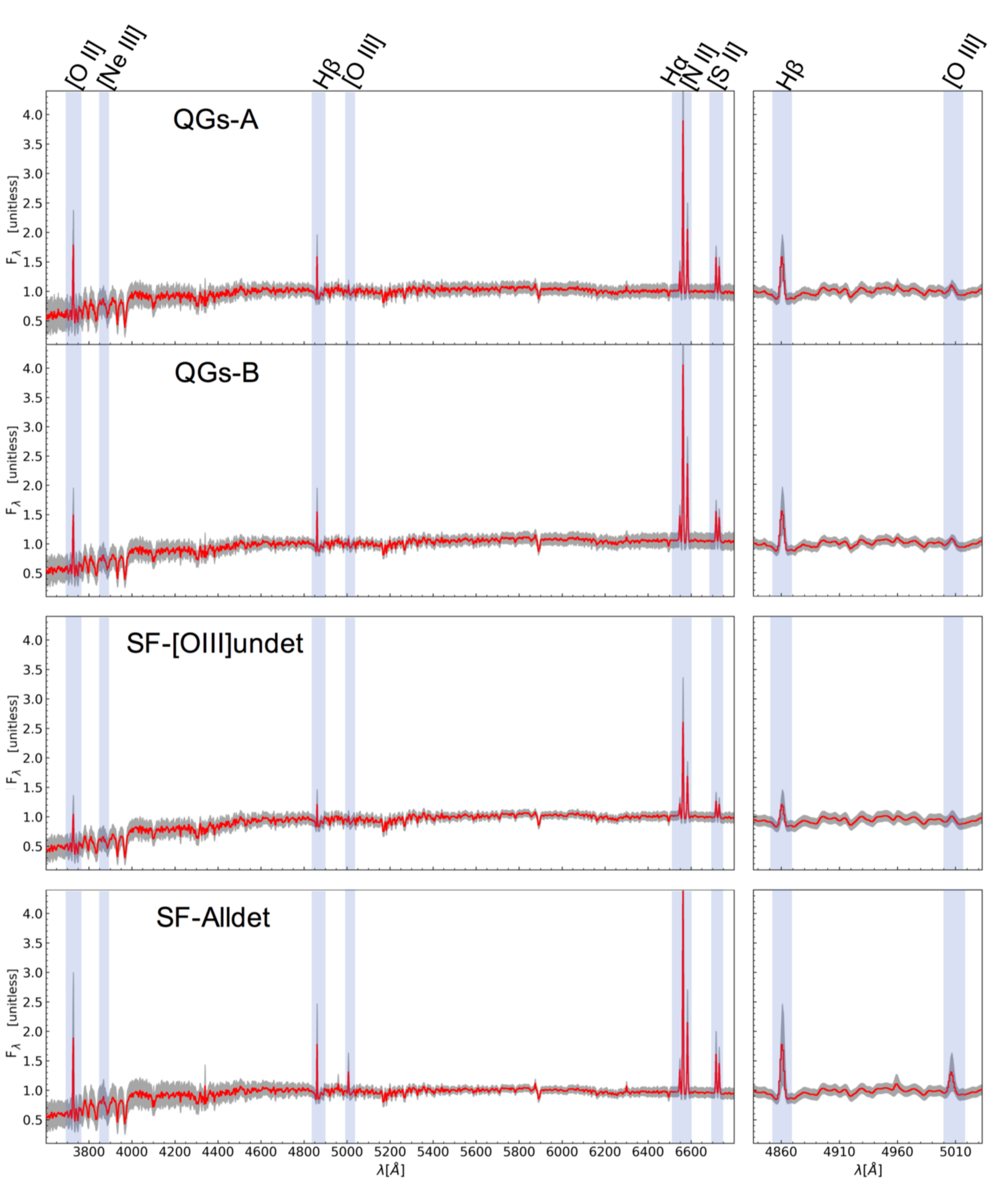}
   \caption{The median stacked spectra (in red) of QGs-A, QGs-B, SF-[\ion{O}{iii}]undet and SF-Alldet galaxies. The gray shaded area represents the dispersion of the stacked spectra. The percentiles of the redshift distributions of galaxies in each stacked spectrum are listed in \autoref{tab:060217_1}.}
         \label{fig:stack_DnC}
\end{figure*}
\begin{figure*}
   \centering
   \includegraphics[trim=0.9cm 0 0 0, clip, width=\linewidth]{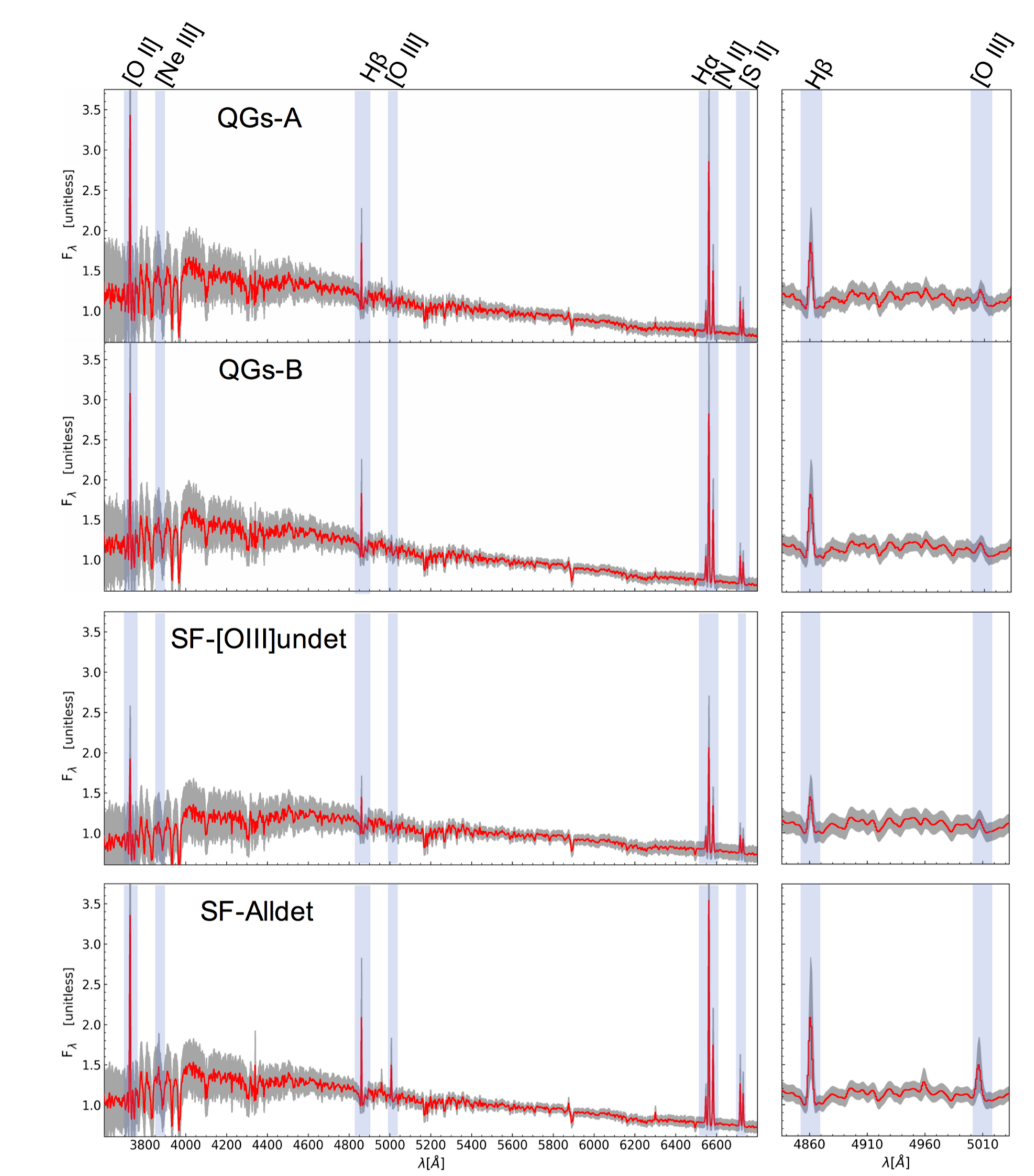}
   \caption{The median stacked spectra corrected for the dust extinction (in red) of QGs-A, QGs-B, SF-[\ion{O}{iii}]undet and SF-Alldet galaxies. The gray shaded area represents the dispersion of the stacked spectra. The percentiles of the redshift distributions of galaxies in each stacked spectrum are listed in \autoref{tab:060217_1}.}
         \label{fig:stack_meth}
\end{figure*}

We first inspect the spectra of our QG candidates. In order to increase their S/N, in particular around [\ion{O}{iii}] and [\ion{Ne}{iii}] to confirm their low ionisation status, we stack their spectra. 
\autoref{fig:stack_DnC} shows the median stacked spectrum of QGs-A and QGs-B. As a comparison, we show also the spectra of two control samples, stacking 
%about 300 galaxies, randomly extracted 
all galaxies from SF-Alldet and SF-[\ion{O}{iii}]undet samples, respectively, in the same mass and redshift range of QGs. 
The [\ion{O}{ii}], H$\alpha$ and H$\beta$ lines (i.e. low-ionisation lines) are the strongest emission lines, while [\ion{O}{iii}] and [\ion{Ne}{iii}], which are high-ionisation emission lines, are very weak in both QGs stacked spectra despite the increased S/N (see the  
zoom of the stacked spectra in the wavelength range around the [\ion{O}{iii}] and the H$\beta$ lines). %, as we expect in our method, confirming the low ionisation level of our QGs. % in \autoref{fig:stack_zoom}, together with a control stacked spectrum of about 300 star-forming galaxies, randomly extracted from our SF-Alldet galaxies at the same QGs mass and redshift range.
%This analysis confirms, therefore, the lack of [\ion{O}{iii}] (and other emission lines of high-ionisation, such as [\ion{Ne}{iii}]) in both QGs stacked spectra, despite the increased S/N. 

Furthermore, in order to measure high-to-low ionisation emission line ratio and confirm the low ionisation level of our QGs, we derive the dust-corrected stacked spectrum, correcting the individual spectra for the dust extinction\footnote{We derive it from the H$\alpha$/H$\beta$ flux ratio using the \cite{Calzetti2000} extinction law, the same that we have adopted for the dust correction of the emission lines.} before stacking them together (see \autoref{fig:stack_meth}). Also in this case we confirm the weakness of [\ion{O}{iii}] (and of other high-ionisation emission lines, such as [\ion{Ne}{iii}]) in both QGs stacked spectra.
We, further, note that in QGs spectra the stellar continuum is blue, suggesting a still young mean stellar population, consistent with a recent quenching of the star formation (see \citetalias{Citro2017} for discussion on the expected colors of QGs).

\begin{table*}
	\centering
	\caption{Main emission lines ratios 
    %and other spectral features 
    measured on stacked spectra of QGs-A, QGs-B, SF-[\ion{O}{iii}]undet and SF-Alldet samples. }
	\label{tab:emlin_meas}
	\begin{tabular}{lcccc} 
	Property & QGs-A & QGs-B & [\ion{O}{iii}]und. & SF-Alldet  \\
	\hline
	\OH		& $-1.25 \pm 0.04$		&$-1.34 \pm 0.04$ &$-1.19 \pm 0.02$	&$-0.92 \pm 0.04$ \\
	\NO			& $-0.23 \pm 0.02$ 		&$-0.29 \pm 0.02$ &$-0.16 \pm 0.02$	&$-0.29 \pm 0.02$\\
	log([\ion{O}{iii}]/H$\beta$)	 		& $-0.79 \pm 0.04$  &$-0.88 \pm 0.04$ &$-0.73 \pm 0.03$ &$-0.47 \pm 0.02$	 \\
	log([\ion{N}{II}]/H$\alpha$)		& $-0.48 \pm 0.02$      &$-0.42 \pm 0.02$ &$-0.44 \pm 0.02$ &$-0.44 \pm 0.04$ 	\\
	log([\ion{O}{iii}]/[\ion{N}{II}])		& $-0.77 \pm 0.03$          &$-0.92 \pm 0.03$ &$-0.75 \pm 0.02$ &$-0.49 \pm 0.02$  \\
	D$_\text{n}$4000 & $1.23 \pm 0.01$ &$1.25 \pm 0.01$ &$1.34 \pm 0.01$ &$1.23 \pm 0.01$ \\
	EW$_\text{rf}$(H$\alpha$) [\AA] & $-18.62 \pm 0.09$ &$-18.19 \pm 0.07 $ &$-12.30 \pm 0.01$ &$-21.38 \pm 0.01$	 \\
	\hline
\end{tabular}		 
\end{table*}

%#########
We run the {\em Gandalf} code \citep{Sarzi2006,Cappellari2004} on the dust-corrected stacked spectra. We fit the continuum with the stellar population synthesis models of \citet{Bruzual2003}, used also by \citet{Tremonti2004}, and measure the main emission lines and spectral properties on the stacked spectra. We list them in \autoref{tab:emlin_meas} and show them in 
\renewcommand{\figureautorefname}{Figures}
\autoref{fig:methodA_summ}, \ref{fig:methodB_summ}, \ref{fig:160117_2} and \ref{fig:160117_9}. 
\renewcommand{\figureautorefname}{Figure}

%We chose to use the best-fit model obtained with Gandalf to perform independent measures of the emission lines on the stacked spectra. 
%Hence, we subtracted the Gandalf best-fit model (stellar continuum) from the stacked spectra and we fitted a Gaussian to the emission lines (H$\beta$, [\ion{O}{iii}] and H$\alpha$) in the residue spectra.   
From this analysis, we find evident H$\alpha$ emission %\sout{still} strong 
in QGs stacked spectra, although slightly weaker than in the star-forming, and given that the samples have similar median redshift, we find that
$\tfrac{\text{L(H}\alpha)_\text{QGs-A}}{\text{L(H}\alpha)_\text{SF-Alldet}} = 0.82 \pm 0.07$, while $\tfrac{\text{L(H}\alpha)_\text{QGs-B}}{\text{L(H}\alpha)_\text{SF-Alldet}} = 0.91 \pm 0.08$.
Further, we measure, in particular, the [\ion{O}{iii}]/H$\alpha$ and [\ion{N}{II}]/[\ion{O}{ii}] ratios of the median stacked spectra (see \autoref{tab:emlin_meas}),
%\sout{log([\ion{O}{iii}]/H$\alpha)_\text{QGs-A} = -1.25$ dex and log([\ion{O}{iii}]/H$\alpha)_\text{QGs-B} = -1.34$ dex}, 
obtaining values consistent with the low ionisation level of our QG candidates and 
far below
%\sout{On the contrary, 
the control sample of star-forming galaxies. %has a higher value of [\ion{O}{iii}]/H$\alpha$, %compatible with ongoing star formation, 
%with log([\ion{O}{iii}]/H$\alpha)_\text{SF} = -0.92$ dex.}   %with respect to the star-forming population one.
We note, instead, that the [\ion{O}{iii}]/H$\alpha$ value measured on the SF-[\ion{O}{iii}]undet stacked spectrum is intermediate between SF and QG candidates, %($-1.19$), 
suggesting that also SF-[\ion{O}{iii}]undet galaxies have a lower ionisation state with respect to the star-forming population, but not as extreme as QG candidates.
%We also calculate the median of the flux ratios measured on the single spectra for the SF-Alldet galaxies, and we obtain %median log([\ion{O}{iii}]/H$\alpha)_\text{QGs-A})= -1.14$ dex, log([\ion{O}{iii}]/H$\alpha)_\text{QGs-B} = -1.29$ dex, which are lower values with respect to that measured on the stacked spectra. 
%Instead, 
%a median log([\ion{O}{iii}]/H$\alpha_\text{SF}) = -0.90$ dex that is in agreement with the ratio measured in the median stacked spectrum.
%}
%{\color{red}
In 
\renewcommand{\figureautorefname}{Figures}
\autoref{fig:methodA_summ}, \ref{fig:methodB_summ} and \ref{fig:160117_2}
\renewcommand{\figureautorefname}{Figure}
we show the ratios measured on stacked spectra. In particular, from 
\renewcommand{\figureautorefname}{Figures}
\autoref{fig:methodA_summ} and \ref{fig:methodB_summ}
\renewcommand{\figureautorefname}{Figure}
we confirm that the ratios measured on stacked spectra of QGs are consistent with both our selection criteria and a low-ionisation state, while the ratios for star-forming galaxies lie consistently on their median relations, compatible with ongoing star formation (see \autoref{fig:121216_1_BIS} and \citetalias{Citro2017}). 
Finally, we note that the ratios for SF-[\ion{O}{iii}]undet lie only slightly above our selection criteria, suggesting a lower ionisation level with respect to SF galaxies.
This also indicates that also amongst these galaxies there are good QG candidates. %, despite the absence of [\ion{O}{iii}] emission line in their single spectra. 
In this case, the limiting flux of the survey does not allow to pre-select individual QG candidates from single spectra, and to separate them from a residual contamination of SF galaxies.  From this analysis, we also confirm that QGs-A have slightly higher [\ion{O}{iii}]/H$\alpha$ and [\ion{O}{iii}]/[\ion{N}{II}] values, and slightly lower [\ion{N}{II}]/[\ion{O}{ii}] and [\ion{N}{II}]/H$\alpha$ values compared to QGs-B, suggesting a lower value for their metallicity, consistently with, on average, lower masses (see \autoref{colmassect}).
We further note that the [\ion{N}{II}]/[\ion{O}{ii}] ratios of our QGs are in both cases similar to those of the SF-Alldet sample, suggesting that our QGs have metallicities similar to the ones of the SF galaxies parent sample.
%In a simple closed box model a metallicity difference smaller than 0.1 dex could be interpreted as a rapid quenching with a time delay of  $\Delta T<1 $ Gyr after strangulation \citep[see][]{Peng2015}.
%}

\subsubsection{D$_\text{n}$4000 vs EW(H$\alpha$)}
\begin{figure}
   \centering
   \includegraphics[width=\columnwidth]{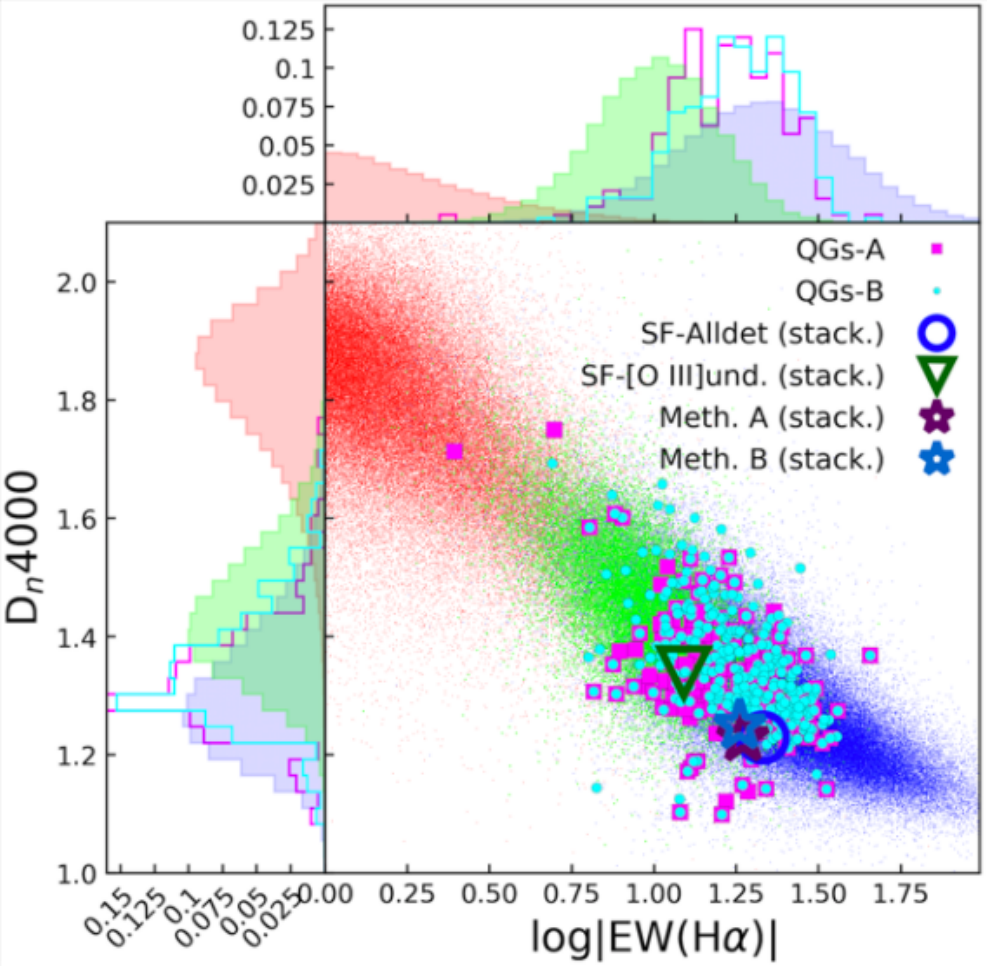} \hfill
    \caption{The D$_\text{n}$4000 as a function of the EW(H$\alpha$). The layout is the same of \autoref{fig:methodB_summ}. }
         \label{fig:160117_9}
\end{figure}
In this subsection we analyse two spectral features of QGs.
The rest-frame equivalent width EW(H$\alpha$), that represents an excellent indicator of the presence of young stellar populations \citep[e.g.][]{Levesque2013} and of the specific SFR (sSFR), and the break at 4000 \AA \; rest-frame, that provides an estimate of the age and metallicity of underlying stellar populations %with mean stellar ages lower than a few Gyr 
\citep[e.g.][]{Moresco2012}.
Using them jointly allows to qualitatively evaluate the connection between the newborn stars and the mean stellar population of the galaxy.

In \autoref{fig:160117_9} we show the relation between the D$_\text{n}$4000 and log($|\text{EW(H}\alpha)|$). Some interesting trends emerge from it. %\autoref{fig:160117_9}. 
As expected, there is a strong anti-correlation between these quantities % with galaxies with older stellar population have a low contribution from young stellar population and a low sSFR. Moreover, 
%both the indicators allow to 
and a clear separation of the no-H$\alpha$ galaxies from the SF ones.
%, and there is a wide intermediate region  around log$|\text{EW(H}\alpha)| \sim 0.5$ \AA\ with low galaxy density. 
%Furthermore, also t} 
Furthermore, we note that the SF-[\ion{O}{iii}]undet sample is shifted  toward higher D$_n$4000 and lower log$|\text{EW(H}\alpha)|$ with respect to the distributions of the SF-Alldet sample.
%{\color{cyan} Indeed, we perform a Kolmogorov - Smirnov test (hereafter KS test) and we confirm, at a significant level $\alpha$ = 0.05, that the D$_n$4000 and EW(H$\alpha$) distributions of the SF-Alldet and SF-[\ion{O}{iii}]undet are different.}
This suggests that the SF-[\ion{O}{iii}]undet galaxies are characterized, on average, by older stellar populations than the SF-Alldet ones.
Finally, we find that both QGs-A and -B lie in a region between the bulk of SF-Alldet and the SF-[\ion{O}{iii}]undet. 
Interestingly, a few QGs show an intermediate EW(H$\alpha$) but very low D$_\text{n}$4000, that could be a fingerprint of recent star-formation quenching.
We confirm these differences by the measurements from the stacked spectra, i.e. %\sout{.also the location of the 
the values of D$_\text{n}$4000 and log($|\text{EW(H}\alpha)|$) of QGs-A and -B
%\sout{median} {\color {cyan} stacked} 
 are intermediate between SF-Alldet %and %[\ion{O}{iii}]undet %\sout{
and no-H$\alpha$ sample
(see values reported in \autoref{tab:060217_1}  and \autoref{fig:160117_9}).
 %Indeed, we perform a Kolmogorov - Smirnov test (hereafter KS test) and we confirm, at a significant level $\alpha$ = 0.05, that the D$_n$4000 and EW(H$\alpha$) distributions of QGs-A and -B are differents from the distributions of the SF-Alldet and SF-[\ion{O}{iii}]undet.
%%{\color {cyan} %\sout{We perform KS tests with the null hypothesis that, at a significant level $\alpha$ = 0.05, the D$_n$4000 distribution of the QGs is the same of both that of SF-Alldet and [\ion{O}{iii}]undet.
%%The test returns p-values of 2.8e-6 and 6.3e-14, respectively for QGs-A and -B appropriately for the SF-Alldet distribution, and p-values of 9.6e-38 and 3.8e-35 with respect to that of [\ion{O}{iii}]undet.
%%Since the p-values are very lower than the significance level chosen, we reject the null hypothesis. 
%We also repeat the KS test for the EW(H)$\alpha$ distributions, 
%%obtaining p-values of 1.6e-11 and 4.8e-14, 
%with respect to the SF-Alldet distribution, and %p-values of 2.7e-38 and 1.1e-74 
%with respect to that of [\ion{O}{iii}]undet and even for EW(H)$\alpha$ we reject the null hypothesis.}
%Indeed%} 
The different distribution of these populations in both D$_n$4000 and EW(H$\alpha$) is also confirmed by Kolmogorov-Smirnov tests (hereafter KS) at high significance level. 

This analysis suggests that our QG candidates have stellar populations which are intermediate between SF and already quenched galaxies, confirming that they are %\sout{stopping} 
interrupting their SF.

%%%%%%%%%%%%%%%%%%%%%%%%%%%%%%%%%%%%%%%%%%%%%%%%%%
%%%%%%%%%%%%%%%%%%%%%%%%%%%%%%%%%%%%%%%%%%%%%%%%%%
%%%%%%%%%%%%%%%%%%%%%%%%%%%%%%%%%%%%%%%%%%%%%%%%%%

%%%%%%%%%%%%%%%%%%%%%%%%%%%%
% Properties of the quenching candidates
%%%%%%%%%%%%%%%%%%%%%%%%%%%%

%\section{Properties of the quenching candidates}
\begin{figure*}
   \centering
   \includegraphics[width=\linewidth]{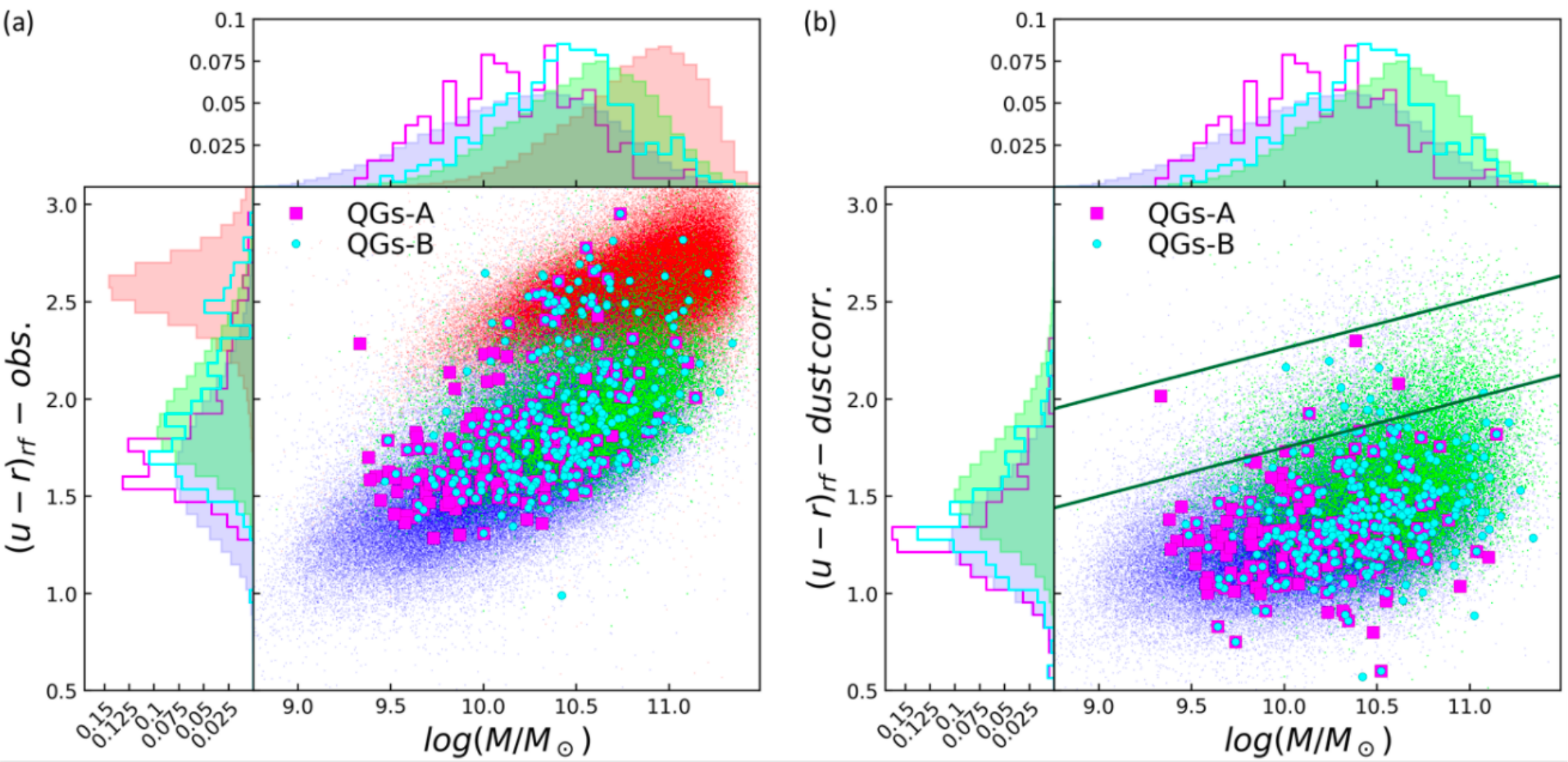} \hfill
    \caption{(a) The rest-frame (u-r) colour - mass diagram. The colours are not corrected for the dust extinction. (b) The same diagram with colours corrected for the dust extinction. 
%About the symbols shape and colour 
We represent the SF-alldet subsample with small blue dots, the SF-[\ion{O}{iii}]undet subsample with small green dots and the control sample of galaxies with no-H$\alpha$ emission with small red dots. We use magenta squares for QGs-A and cyan circles for QGs-B.
The two dark green straight lines represent the edge of the green valley defined by \protect\cite{Schawinski2014}.}
         \label{fig:160117_3}
\end{figure*}

%In this section we analyse the %fundamentals 
%properties of the QGs in order to identify or to constrain plausible quenching mechanism. 

%In \autoref{tab:060217_1} we report the parameters distribution of the candidates, compared against three control samples: SF-Alldet, SF- [\ion{O}{iii}]undet (the subsample in which the candidates are selected) and no-H$\alpha$. 

\subsection{QGs in the colour-mass diagram}
\label{colmassect}
\begin{figure}
   \centering
   \includegraphics[width=\columnwidth]{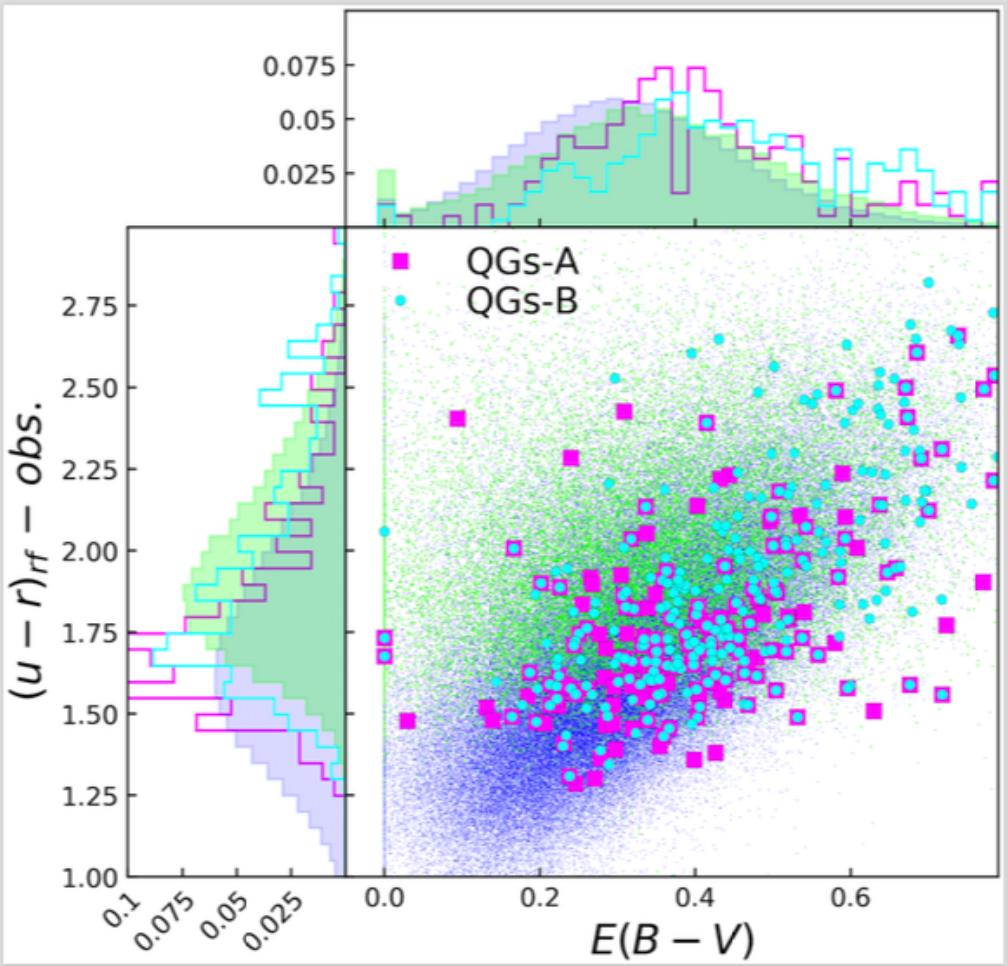} \hfill
    \caption{The (u-r) colours as a function of the  E(B-V). The layout is the same of \autoref{fig:160117_3}.}
         \label{fig:160117_5}
\end{figure}
\begin{figure}
   \centering
   \includegraphics[width=\columnwidth]{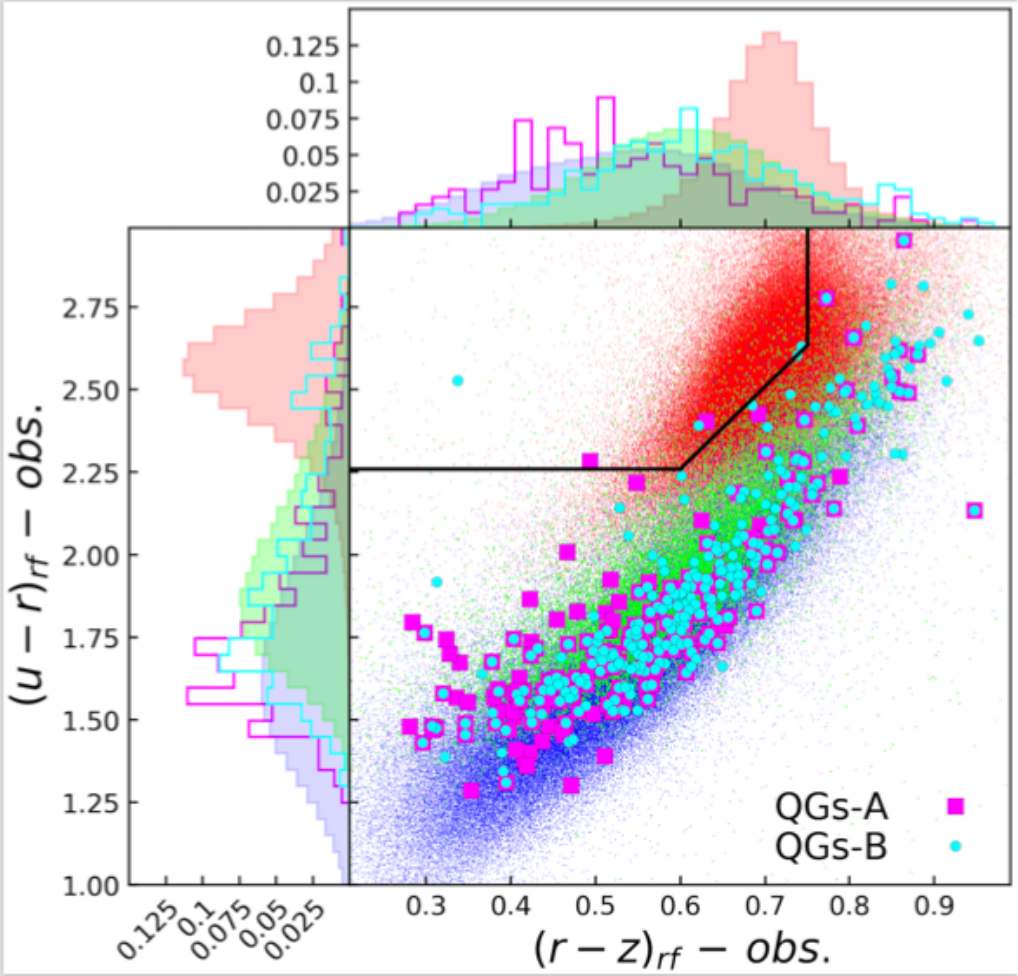} \hfill
    \caption{The rest-frame dust-uncorrected colours (u-r) - (r-z) diagram. The layout is the same of \autoref{fig:160117_3}. The boundary in black is from \protect\cite{Holden2012}. }
         \label{fig:color_color_1}
\end{figure}

\autoref{fig:160117_3}~\textcolor{blue}{(a)} shows the rest-frame, dust-uncorrected colour (u-r) as a function of stellar mass. 
Our SF-alldet sample forms the well-known blue cloud, while the complementary sample of no-H$\alpha$ emission sample shapes the red sequence. The SF-[\ion{O}{iii}]undet sample overlaps with the blue cloud in the intermediate region between the two sequences. 
The QGs-A are mainly located in the blue cloud region at colours $1.5 \lesssim \text{(u-r)} \lesssim 2.1$, while only a few of them have redder colours, near or in the lower part of the red sequence.
We verify that the red colours of these QGs are due to a strong dust extinction (see following discussion and  \autoref{fig:160117_5}).
The colours of QGs-B have instead a larger spread ($1.6 \lesssim \text{(u-r)} \lesssim 2.4$), with a median value redder than the SF-Alldet sample and QGs-A, but still blue. Their colour distribution presents also a significant tail reaching the red sequence ($15.9\%$ of the sample has colours (u-r) $>  2.4$). 
As for QGs-A, we verify that these red candidates are reddened by dust extinction (see \autoref{fig:160117_5}).

In \autoref{fig:160117_5} we show the observed \text{(u-r)} colour as a function of the E(B-V) derived from the H$\alpha$/H$\beta$ ratio, in order to analyse the contribution of the dust extinction to the colour distribution. 
  Obviously, the no-H$\alpha$ control sample is not included, since its galaxies have S/N(H$\alpha$) and S/N(H$\beta$) lower than 3. 
There is a clear correlation between colour and E(B-V), also in QGs samples, with the reddest galaxies having the highest values of E(B-V).
About 16\% of QGs-A %and B 
show E(B-V) higher than 0.6, while the same percentage of QGs-B have even higher dust extinction, showing E(B-V)$>0.7$. %and 0.7, respectively
As anticipated, the QGs with the reddest colours are those with the highest E(B-V) values, which confirm that their intrinsic colours are still blue, as expected from their recent quenching phase \citepalias[see also the discussion in][]{Citro2017}.

In order to better distinguish the dust-reddened quenching candidates from the intrinsic red ones, we exploit the \cite{Holden2012} rest-frame dust-uncorrected colour-colour plane (u-r) vs (r-z), showing the results in \autoref{fig:color_color_1}. 
Only a few candidates are located in the region of pure passive red galaxies, whose boundaries are defined by \cite{Holden2012}. 
On the contrary, the other red candidates are actually reddened by the dust extinction.

Finally, \autoref{fig:160117_3}~\textcolor{blue}{(b)} shows the colour-mass diagram with the (u-r) corrected for the dust extinction. 
In particular, we adopt the attenuation law of \cite{Calzetti2000}, with the stellar continuum colour excess E$_\text{S}$(B-V) = 0.44 E(B-V).
As already shown, we confirm that none of our QG candidates has intrinsic red colours and only a few of them lie in the green valley region defined by \cite{Schawinski2014}.
However, although they are mainly in the blue cloud, their colour distributions are different from that of SF-Alldet, showing a peak at (u-r) $\sim$ 1.3 and on average redder colors (see \autoref{fig:160117_3}~{\color {blue} (b)}). This is also confirmed by the KS test at a significance level $\alpha$ = 0.05.

We stress here that most of our QG candidates would not be selected using dust-corrected green colours, i.e. they do not lie within the so called ``Green Valley'', which separate star-forming galaxies from quiescent passive ones.

From our analysis we find that the mass distribution of QGs-A is spread (i.e. 16th-84th percentiles) over the range $9.8 \lesssim \text{log(M}/\text{M}_\odot\text{)} \lesssim10.6$, being comparable with that of SF-Alldet galaxies ($9.7 \lesssim \text{log(M}/\text{M}_\odot\text{)} \lesssim10.7$), however, QGs-A are slightly less massive than the global SF-[\ion{O}{iii}]undet sample ($10.1 \lesssim \text{log(M}/\text{M}_\odot\text{)} \lesssim10.9$). 
The masses of QGs-B are in the range $\sim10 < \text{log(M}/\text{M}_\odot\text{)} < \sim10.8$, i.e. they are more massive than those derived by method A.
This evidences are also confirmed by the KS test at a significance level $\alpha$ = 0.05, verifying that the masses of QGs-A and SF-Alldet are drawn from the same distribution (i.e. \emph{p}-value = 0.052) , differently from that of QGs-B. Moreover, both the QGs masses have distributions which are different from that of [\ion{O}{iii}]undet.
Furthermore, we note that no QG candidates have masses lower than $\text{log(M}/\text{M}_\odot\text{)} < \sim9.5$ for both methods (A and B). This suggest that, as expected in a downsizing scenario, quenching has not started yet for the low-mass galaxies. This is supported also by the lack of a population of low-mass red galaxies among our no-H$\alpha$ sample in the red sequence.

%########
\subsection{Star formation rates}
\begin{figure}
   \centering
   \includegraphics[width=\columnwidth]{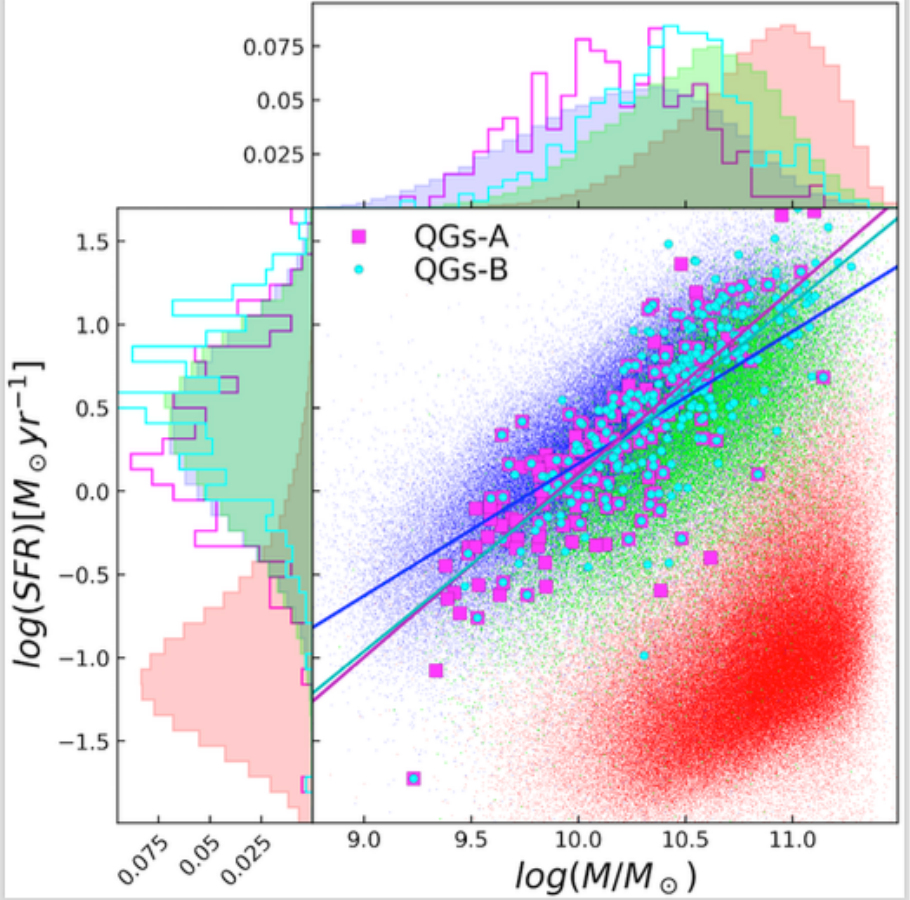} \hfill
    \caption{The log(SFR$_\text{tot}$) as a function of the stellar mass.  The layout is the same of \autoref{fig:160117_3}.The blue straight line represents the main-sequence (MS) of the SF-Alldet sample, while the magenta and the cyan ones are the MS of QGs-A and QGs-B, respectively.}
 \label{fig:160117_4}
\end{figure}

\autoref{fig:160117_4} shows the SFR-mass plane of our samples. 
As described in \autoref{subsec:sampsel}, the SFR is estimated from the dust-corrected H$\alpha$ luminosity. For the no-H$\alpha$ sample, instead, the SFR are derived from a multi-band photometric fitting.
We stress, however, that for the QG candidates these SFRs estimates should be considered as upper limits to their current SFR, due to their past SFR preceding the quenching.
Indeed, even when the O stars die, the longer-lived B stars have sufficient photons harder than 912 \AA\ to ionise hydrogen, explaining their H$\alpha$ emission.
In this case their H$\alpha$ emission can be considered as an upper limit to their current SFR.
As expected, our SF-Alldet sample forms the well known SF main sequence\footnote{The straight-line representing the SF-Alldet MS is log(SFR) = $0.79 \times \text{log(M/M}_\odot) -7.74$.} (MS), while the SF-[\ion{O}{iii}]undet sample lies just below it, but it is well separated and above the  no-H$\alpha$ sample of low-SFR/passive galaxies. 
We find that both QGs-A and -B show high SFRs, with only few of them having very low SFRs, compatible with them being already passive. The QGs-A sample has SFR in the range $0.6 < \text{SFR}$ [M$_\odot$ yr$^{-1}] < 11.5$, with a distribution similar to that of the SF-Alldet galaxies\footnote{The straight-lines representing the QGs MS are log(SFR) = $1.10 \times \text{log(M/M}_\odot) -10.93$ and log(SFR) = $1.04 \times \text{log(M/M}_\odot) -10.36$, respectively for QGs-A and -B.}(see \autoref{tab:060217_1}), but above the SF-[\ion{O}{iii}]undet galaxies. These evidences are confirmed by KS tests with a significance level $\alpha = 0.05$. 
Instead, the KS tests show that the SFR distribution of QGs-B and SF-Alldet are different. This result arises also from the 16th-84th percentiles of the distribution (see \autoref{tab:060217_1}), where the SFR of QGs-B are slightly shifted towards higher SFR than those of the SF-Alldet population. This effect is mainly due to the higher average mass for QGs-B sample. 

From this analysis we also stress that most of our QG candidates would not be selected as intermediate between SF and passive quiescent ones from the SFR-mass plane.

%\begin{landscape}
\renewcommand{\arraystretch}{1.4}
\begin{table*}
	\centering
	\caption{Fundamental properties of the QGs, compared against three control samples: SF-Alldet, SF- [\ion{O}{iii}]undet (the subsample in which the candidates are selected) and no-H$\alpha$. For each parameter we report the 50th (16th, 84th) percentiles of its distribution.}
	\label{tab:060217_1}
	\begin{tabular}{lccccc} 
Property & QGs-A  &  QGs-B & SF-[\ion{O}{iii}]undet & SF-Alldet & no-H$\alpha$\\
\hline
N. &     192 &     308 &   25911 &  148145 &  201527 \\
redshift &  0.08  (0.05, 0.11) &  0.08  (0.06, 0.13) & 0.12  (0.07, 0.16) &  0.08  (0.05, 0.13) &  0.12 (0.08, 0.16) \\
%{\color {red} mean z} & {\color {red}0.083} & {\color {red}0.096} & {\color {red}0.113} & {\color {red}0.088} & {\color {red}0.119} \\
log(M/M$_\odot$) & 10.1 ( 9.7, 10.6) & 10.4 (10.1, 10.8) & 10.6 (10.1, 10.9) & 10.2 ( 9.7, 10.7) & 10.9 (10.4, 11.2) \\
(u-r)$_\text{rf}$ obs. &  1.73 ( 1.53,  2.10) &  1.88 ( 1.61,  2.40) &  1.93 ( 1.69,  2.27) &  1.68 ( 1.38,  2.05) &  2.57 ( 2.37,  2.75) \\
E(B-V) &  0.40 ( 0.26,  0.59) &  0.45 ( 0.30,  0.68) &  0.34 ( 0.19,  0.51) &  0.31 ( 0.17,  0.46) & / \\
SFR [M$_\odot$ yr$^{-1}$] &  0.33 ( 0.08,  1.63) &  0.78 ( 0.22,  2.88) &  0.38 ( 0.09,  1.31) &  0.47 ( 0.12,  1.75) &  0.03 ( 0.01,  0.08) \\
log(sSFR) [yr$^{-1}$] & -10.6 (-11.0, -10.3) & -10.5 (-10.8, -10.2) & -11.0 (-11.4, -10.6) & -10.5 (-10.9, -10.1) & -12.4 (-12.7, -11.9) \\
C(R90/R50) &  2.20 ( 2.01,  2.53) &  2.28 ( 2.05,  2.62) &  2.19 ( 1.97,  2.55) &  2.27 ( 2.04,  2.60) &  2.86 ( 2.52,  3.13) \\
D$_n$4000 &  1.31 ( 1.25,  1.41) &  1.34 ( 1.27,  1.45) &  1.43 ( 1.33,  1.55) &  1.30 ( 1.20,  1.42) &  1.84 ( 1.70,  1.95) \\
EW$_\text{rf}$(H$\alpha$) [\AA] & -16.9 (-24.4, -11.6) & -18.1 (-26.1, -12.0) & -10.1 (-15.3, -6.5) & -21.2 (-37.8, -11.8) & -0.8 (-2.1, -0.2) \\
\hline
\end{tabular}		 
\end{table*}
\renewcommand{\arraystretch}{1}

%%%%%%%%%%%%%%%%%%%%%%%%%%%%%%%%%%%%%%%%%%%%%%%%%%
%%%%%%%%%%%%%%%%%%%%%%%%%%%%%%%%%%%%%%%%%%%%%%%%%%
%%%%%%%%%%%%%%%%%%%%%%%%%%%%%%%%%%%%%%%%%%%%%%%%%%
\subsection{Morphology}
\begin{figure}
   \centering
   \includegraphics[width=\columnwidth]{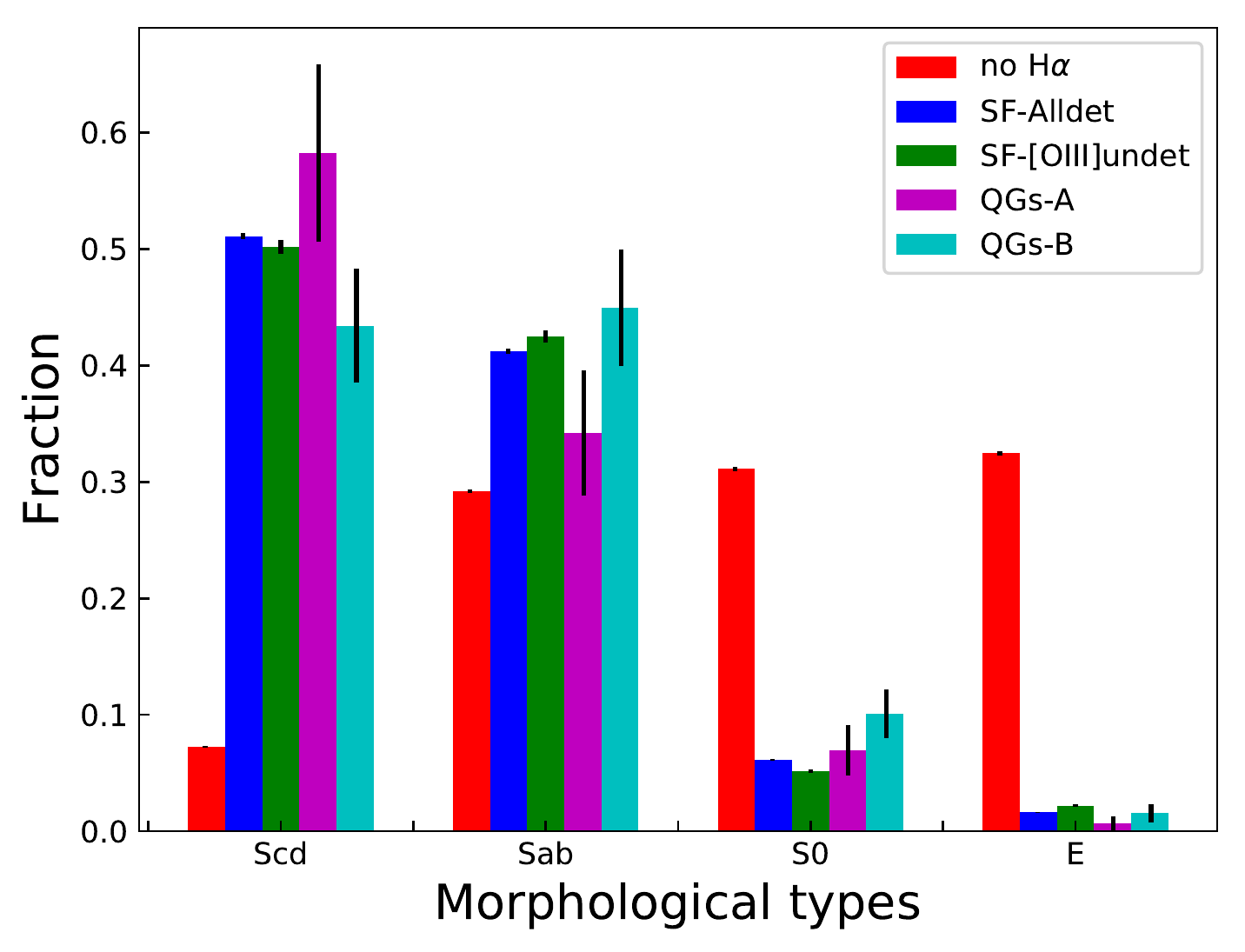} \hfill
     \caption{Distribution of no-H$\alpha$ (red), SF-Alldet (blue), SF-[\ion{O}{iii}]undet (green),  QGs-A (magenta) and QGs-B (cyan) in 4 morphological types (Scd, Sab, S0 and E).}
         \label{fig:morph_1}
\end{figure}
In this section we analyse the morphologies of our quenching candidates.
The favorite scenario for the transformation of star forming galaxies into passive ones suggests both the migration from the blue cloud to the red sequence and the morphological transformation from disks to spheroids \citep[e.g.][]{Faber2007, Tacchella2015}. 
It is still unclear if this transformation occurs during the migration or via dry merging, when a galaxy has already reached the red sequence.  
%Actually, the intermediate situation is more plausible.
%Therefore, 
Our QGs samples, catching the galaxies in an early phase after SF quenching, are therefore crucial to address this open question. 
%We have adopted a statistical approach, investigating the morphological probability distribution of the galaxies from \cite[][]{Huertas-Company2011}\footnote{Downloaded together with the environmental catalogue of \cite{Tempel2014} used in the previous section.}

In \autoref{tab:morph} we report the 16th-50th-84th percentiles of the morphological probability distribution of the four morphological classes (Scd, Sab, S0, E) for %the QGs-A and QGs-B, SF-Alldet, SF-[\ion{O}{iii}]undet  and no-H$\alpha$ 
our subsamples.
In \autoref{fig:morph_1} we further show the distribution built assigning to each galaxy the morphological class with the highest probability.
SF-Alldet and SF-[\ion{O}{iii}]undet galaxies have the same distribution: roughly 50\% of SF objects are late Scd galaxies, while 40\% of them are Sab and less than 10\% are S0. On the contrary, the no-H$\alpha$ sample shows a different distribution, in which the early type classes are more common then the late type ones.
For comparison, the bulk of QGs-A are Scd ($\sim$ 60\%), while 35\% are Sab. Also QGs-B are disk galaxies with a similar probability of being disk dominated Scd galaxies or bulge dominated Sab disk galaxies. Only $\sim 7\%$ of QGs-A and $\sim 10\%$ of QGs-B are instead S0 or E galaxies. 
Therefore, we conclude that our candidates have the same morphology classes as the SF galaxies. 
Therefore, our analysis suggests that no morphological transformation has yet occurred in the early phase after the quenching of the SF.

\begin{figure}
   \centering
    \includegraphics[width=\columnwidth]{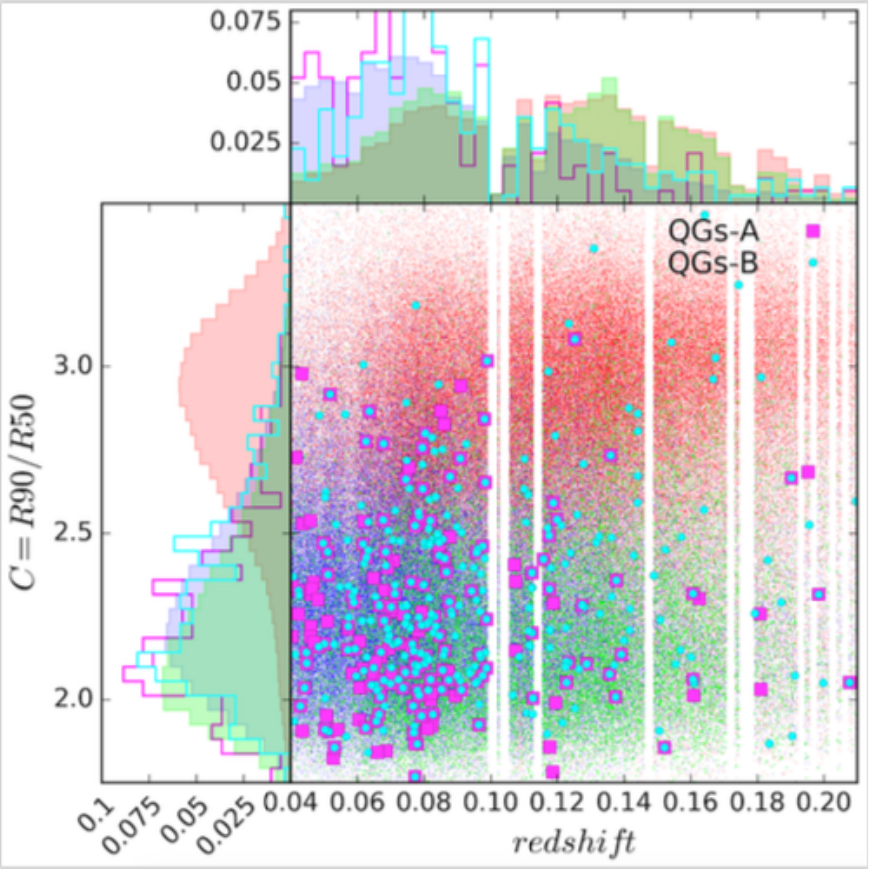} \hfill
    \caption{The light concentration - redshift relation for our sample and QGs. The layout is the same of \autoref{fig:160117_3}}
         \label{fig:C_z}
\end{figure}

We further analyse the concentration-redshift relation, shown in \autoref{fig:C_z}.
The concentration is defined as C=R90/R50, where R90 and R50 are the radii containing 90 and 50 per cent of the Petrosian flux in \emph{r}-band. 
This parameter is strongly linked to the morphology of the galaxies, and there is general consensus that C=2.6 is the threshold concentration dividing early type galaxies from the other types \citep[e.g.][]{Strateva2001}. This value is, indeed, confirmed by the crossing point between the C distributions of our SF and no-H$\alpha$ samples. 
The bulk of QGs have C $< 2.6$, but some of them ($\sim 12\%$ of QGs-A and $\sim 17\%$ of QGs-B) have higher concentrations.
This suggests that they could be quenching galaxies which have experienced morphological transformation during the transition from blue cloud to red sequence.

\begin{table*}
\centering
\caption{The 50th (16th, 84th) percentiles of the morphological probability distribution of QGs-A and QGs-B, compared against three control samples: SF-Alldet, SF-[\ion{O}{iii}]undet (the subsample in which the QGs are selected) and no-H$\alpha$. }
\label{tab:morph}
\begin{tabular}{lcccc}
& P(Scd) & P(Sab) & P(S0) & P(E) \\
\hline
QGs-A 		           &  0.35 (0.13,0.70) &  0.30 (0.17,0.60)   &  0.04 (0.01,0.14) &  0.01 (0.00,0.03)  \\
QGs-B. 		           &  0.25 (0.12,0.65) &  0.35 (0.18,0.65)   &  0.05 (0.02,0.22) &  0.01 (0.01,0.04)  \\
SF-Alldet 		       &  0.33 (0.14,0.65) &  0.35 (0.19,0.61)   &  0.06 (0.03,0.20) &  0.01 (0.01,0.04)  \\
SF-[\ion{O}{iii}]undet &  0.27 (0.07,0.66) &  0.34 (0.15,0.67)   &  0.04 (0.01,0.14) &  0.01 (0.00,0.03)  \\
no-H$\alpha$           &  0.06 (0.03,0.23) &  0.13 (0.04,0.58)   &  0.19 (0.09,0.61) &  0.05 (0.01,0.72)  \\
\hline
\end{tabular}		 
\end{table*}
%%%%%%%%%%%%%%%%%%%%%%%%%%%%%%%%%%%%%%%%%%%%%%%%%%
%%%%%%%%%%%%%%%%%%%%%%%%%%%%%%%%%%%%%%%%%%%%%%%%%%
%%%%%%%%%%%%%%%%%%%%%%%%%%%%%%%%%%%%%%%%%%%%%%%%%%

%%%%%%%%%%%%%%%%%%%%%%%%%%%%
% Environment
%%%%%%%%%%%%%%%%%%%%%%%%%%%%

\subsection{Environment}
In this section, we examine the environment of our sample of QGs. 
Studying the local environment of a galaxy is crucial to disentangle between several known mechanisms able to remove the cool gas needed for star formation. 
\begin{figure}
   \centering
   \includegraphics[width=\columnwidth]{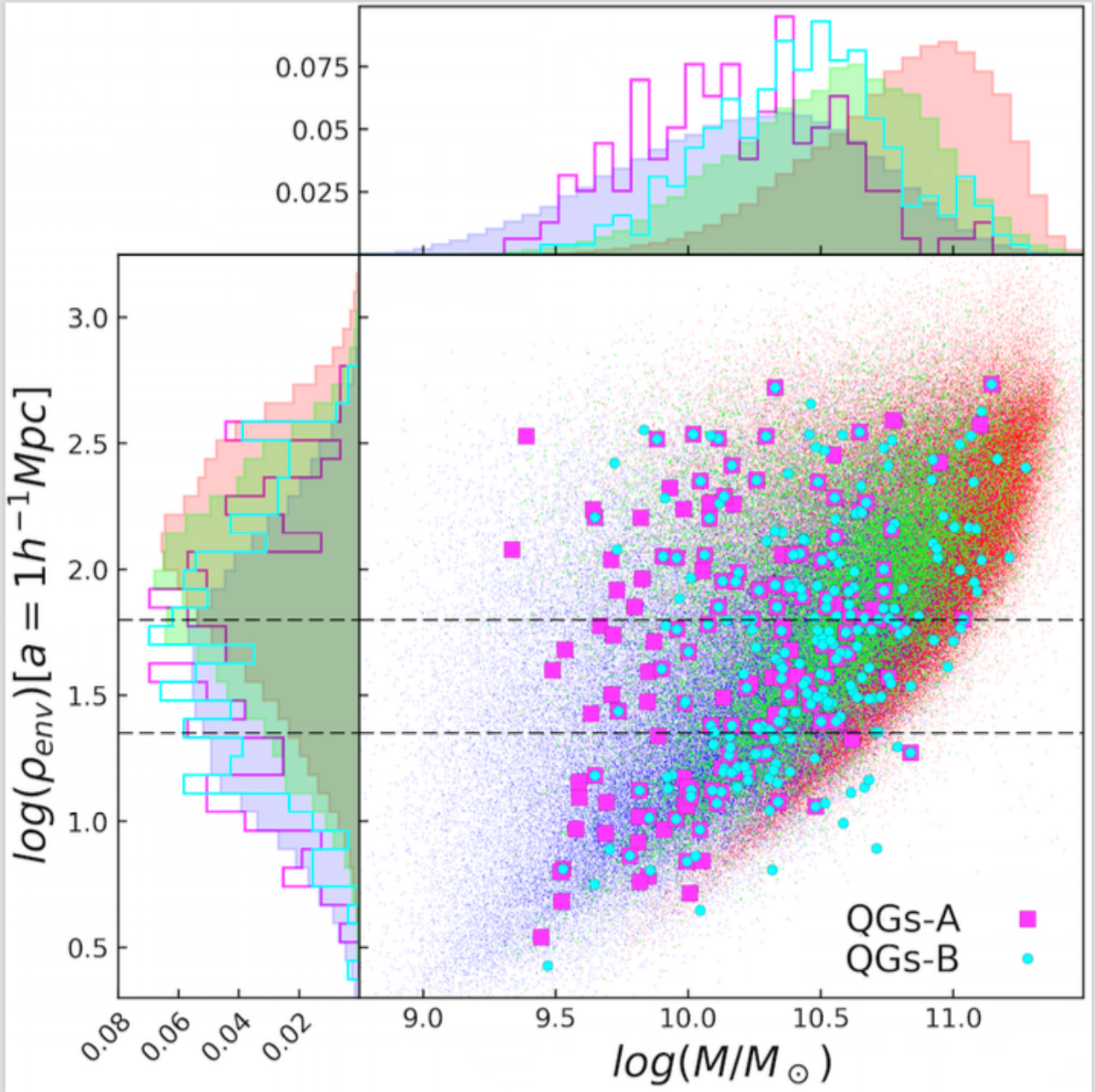} \hfill
    \caption{Normalised environmental density of galaxies $\rho_{\text{env}}$ (smoothing scales of 1 h$^{-1}$ Mpc) vs. stellar mass. The two dashed lines (at $\rho_{\text{env}} = 22.45$ and $\rho_{\text{env}} = 62.87$, respectively) divide the SF-Alldet $\rho_{\text{env}}$ distribution into three tertiles.}
         \label{fig:env_mass}
\end{figure}

\autoref{fig:env_mass} shows the 'environmental density' of galaxies $\rho_{\text{env}}$ normalised at a smoothing radius of 1 h$^{-1}$ Mpc (see \autoref{subsec:sampsel}), as a function of stellar mass.
It is possible to note a general trend, but with a wide spread, in which the highest stellar mass of the galaxies increases for increasing density and this behavior is true also for our QGs. 
We note, in particular, that at the highest densities there are QGs with a large mass spread, while low density environments are populated only by galaxies and QGs with stellar mass lower than $\sim 10^{10.5} \text{M}_\odot$.
Viceversa, galaxies and QGs with the highest masses reside only in high density environments.

\begin{figure}
   \centering
   \includegraphics[width=\columnwidth]{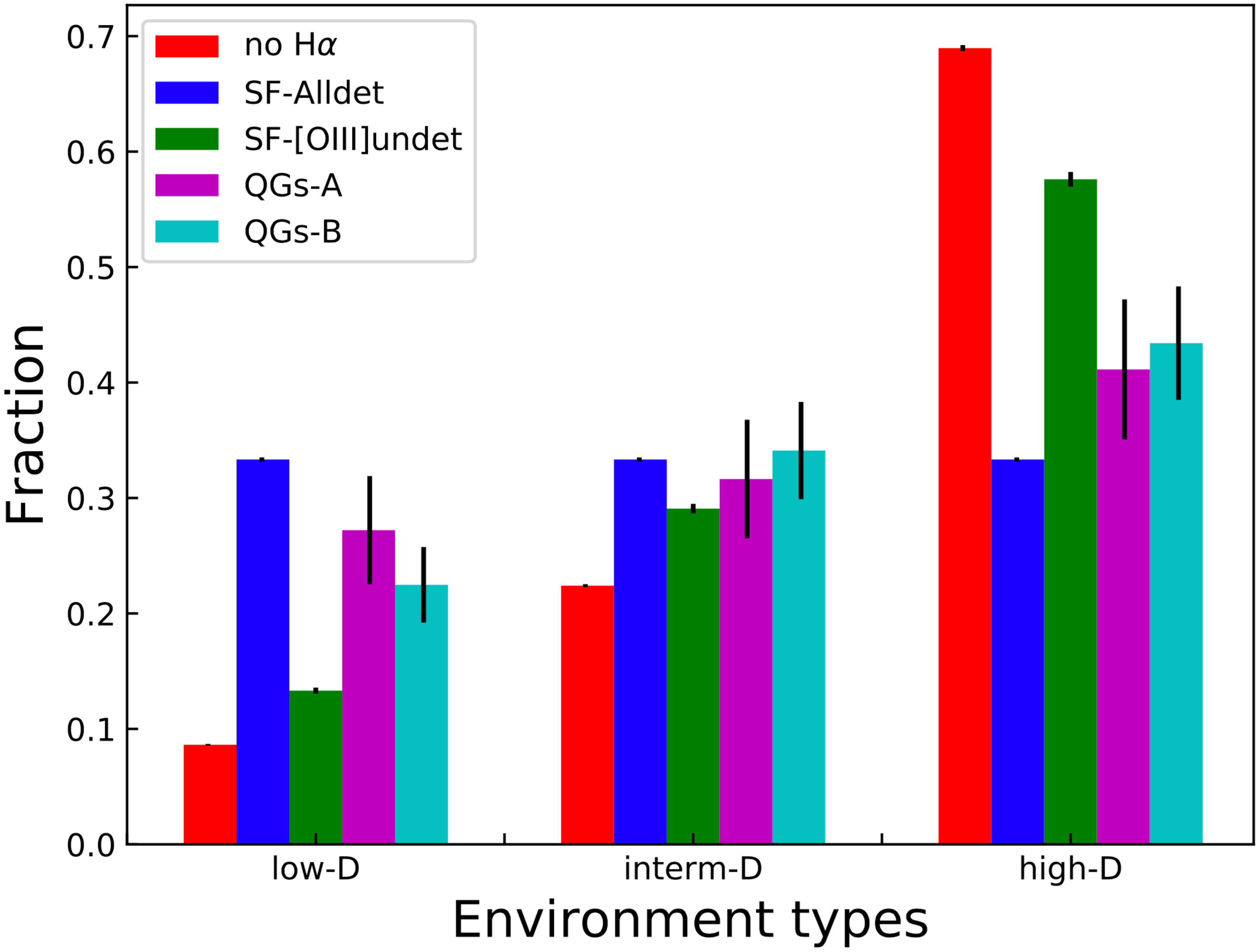} \hfill
    \caption{Distribution of no-H$\alpha$ (red), SF-Alldet (blue), SF-[\ion{O}{iii}]undet (green),  QGs-A (magenta) and QGs-B (cyan) in 3 environment types (i.e. low-D, interm-D, high-D).}
         \label{fig:env_hist_p}
\end{figure}

We divide the sample into three environmental classes, separated at $\rho_{\text{env}} = 22.45$ and $\rho_{\text{env}} = 62.87$, respectively, which are defined basing on the tertiles of the $\rho_\text{env}$ distribution of the parent SF-Alldet population.
We define 'low-D' those galaxies belonging to the first tertile; %(18.79\% of the total galaxy sample)
'interm-D' those in the second tertile %(27.21\%) 
and 'high-D' %(53.99\%) 
those galaxies belonging to the third tertile. 
We compare the environment of the QG candidates against that of the parent SF population, finding a hint of a lack of QGs in low-D environment and an excess in high-D environment, at high significance level ($\sim 3\sigma$) only for QGs-B (see \autoref{fig:env_hist_p}). 
Indeed, a KS test confirms this behavior at a significance level $\alpha$=0.05 only for QGs-B, while  QG-A and SF-Alldet populations appear to have a compatible $\rho_{\text{env}}$.

\begin{table*}
	\centering
	\caption{The environment of the quenching candidates with methods A and B compared against three control samples: SF-Alldet, SF-[\ion{O}{iii}]undet (the subsample in which the candidates are selected) and no-H$\alpha$. The column Global lists the ratio between the number of objects in common with the \protect\cite{ Tempel2014} sample over the total number in each sub-sample.
    }
	\label{tab:060217_2}
	\begin{tabular}{lcccc} 
	& Global & low-D & interm-D  & high-D \\
	&&($\rho_\text{env.}<22.45$) & ($22.45\leq\rho_\text{env.}<62.87$) & ($\rho_\text{env.}\geq62.87$) \\
	\hline
	QGs-A		& 158/192 		&$27.2 \pm 4.7 \%	$	& $31.7 \pm 5.1 \%$		&$ 41.1 \pm 6.1 \%$	\\
	QGs-B		& 258/308 		& $22.5 \pm 3.3 \%$		& $34.1 \pm 4.2 \%$		& $43.4 \pm 4.9 \%$ \\
	SF-Alldet 		& 130651/148145 	&$33.3 \pm 0.2 \%$ 	& $33.3 \pm 0.2 \%$		& $33.3 \pm 0.2 \%$	\\
	SF-[\ion{O}{iii}]undet	& 22667/25911 	& $13.3 \pm 0.3 \%$		& $29.1 \pm 0.4 \%$	&$57.6 \pm 0.6 \%$ 	 \\
	no-H$\alpha$ 	& 174741/201527 	& $8.6 \pm 0.1 \%$	& $22.4 \pm 0.1 \%$ & $69.0 \pm 0.3 \%$ \\
	\hline
\end{tabular}		 
\end{table*}

In \autoref{tab:060217_2} we compare also the fraction in the three different environments of the QG candidates against other two reference samples of no-H$\alpha$ and SF-[\ion{O}{iii}]undet galaxies. 
As it is possible to note, SF-OIIIundet galaxies are even more extreme  than QGs-B, given that the 57.6 $\pm 0.6$\% of them in the high-D tertile, and reside at each mass in environments which are intermediate between SF and no-H$\alpha$ galaxies.

\begin{figure}
   \centering
   \includegraphics[width=\columnwidth]{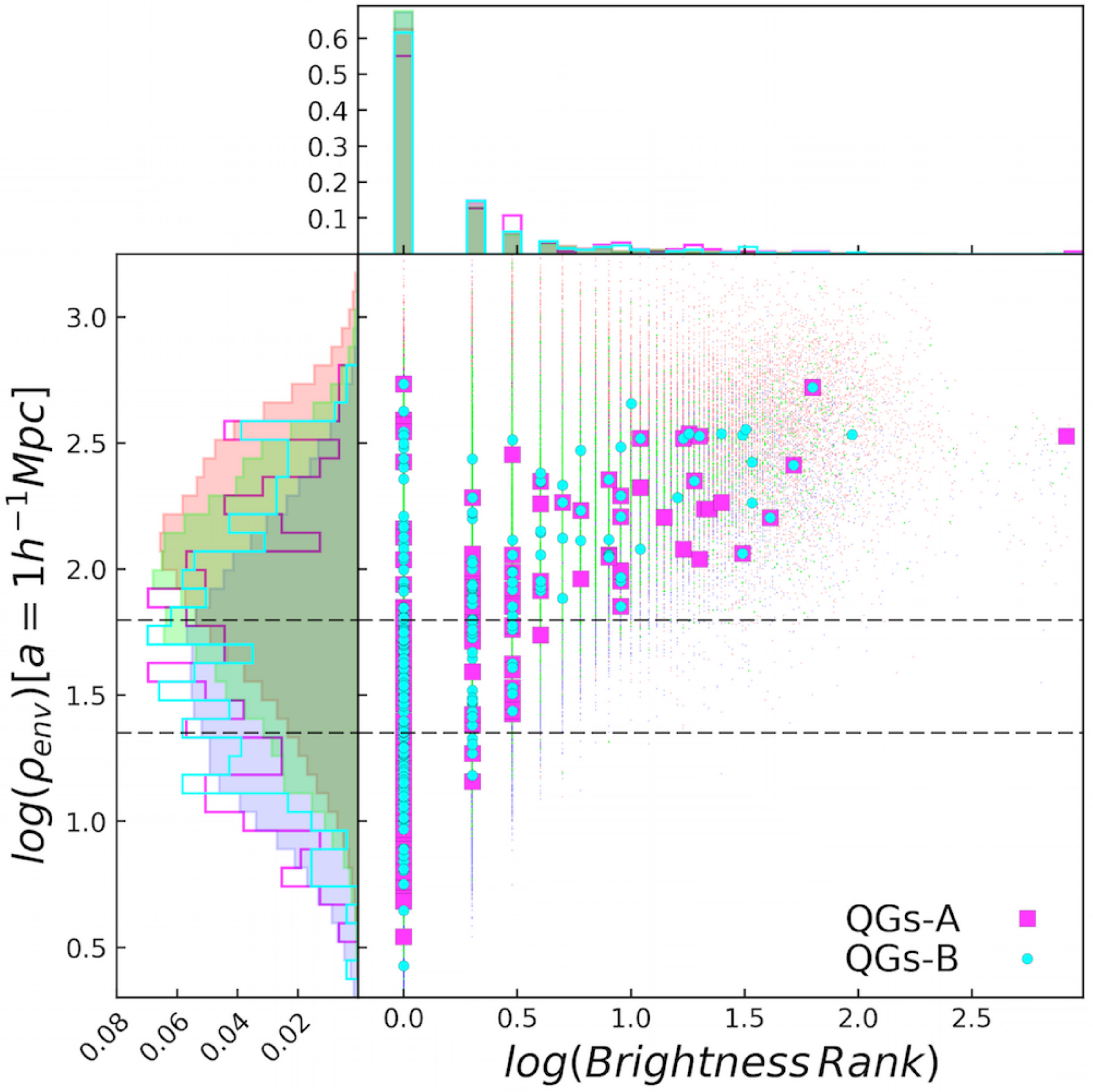} \hfill
    \caption{Normalised environmental density of galaxies $\rho_{\text{env}}$ (smoothing scales of 1 h$^{-1}$ Mpc) vs. the brightness rank. The two dashed lines (at $\rho_{\text{env}} = 22.45$ and $\rho_{\text{env}} = 62.87$, respectively) divide the SF-Alldet $\rho_{\text{env}}$ distribution into three tertiles.}
         \label{fig:env_1}
\end{figure}

We also analyse the Richness (R) and the 'Brightness Rank' (hereafter BR) to evaluate whether the candidates are either the dominant/brightest galaxies or satellites within their group/cluster. BR ranges from the values of the group/cluster richness R to 1.
In particular, BR=richness and BR=1 indicate that the considered galaxy is the faintest or the brightest (and thus the most massive) in its group/cluster, respectively. We define as 'central' a galaxy whose brightness rank is equal to 1.

\autoref{fig:env_1} shows %the %'environmental density' of galaxies
$\rho_{\text{env}}$ %normalised at a smoothing radius of 1 h$^{-1}$ Mpc}, 
as a function of BR of the galaxies in their own environment. 
%(BR) of the galaxies in its own environment (i.e. an index of the luminosity level of the galaxy in the group/cluster it belongs to, where BR = 1 means the most luminous galaxy of the group/cluster while BR = Richness means the least luminous one).
%We evaluated whether the candidates are either the dominant/brightest galaxies or satellites within their group/cluster. 
%Basically, we define 'central' a galaxy whose brightness rank is equal to 1.
%(i.e. the brightest galaxy in its own environment and consequently the most massive one) and 'satellites' the others. 
We find (see \autoref{fig:env_1}) that the bulk of QGs in high-D enviroments are satellites (76.9\% and 63.4\%, respectively for QGs-A and -B), with percentages higher than those of the parent SF-Alldet population (57.3\%) and of the SF-OIIIundet population (46.1\%).
Finally, almost all ($>$ 90\%) galaxies belonging to groups/clusters including more than 30 members (R > 30) are in high-D environment and all the QGs in these extreme dense environments (i.e. $5.70\pm1.95\%$ of QGs-A and $4.65\pm1.37\%$ of QGs-B) are satellites.

%\autoref{fig:070217_2} shows the 'luminosity rank' of the galaxies as a function of the group's richness (Rank $= 1$ means the most luminous galaxy of the groups and Rank = Richness means the least luminous). 
%There is a dichotomy: in small groups (R < 5-10) the candidates do not show preferences and roughly occupy all ranks; instead, in richer groups the candidates have Rank $\sim$ Richness. 
%The bulk of QGs belong to the 50\% of the faintest galaxies in each group; instead, there are only few cases of candidates belonging to the 50\% of the brightest galaxies. 

Therefore, we conclude that our QGs are preferentially satellite galaxies within groups of medium and high density, showing an excess in high density environments compared to SF galaxies.

\section{Discussion}
\subsection{Quenching timescale}
In this section, we use the fraction of our selected QGs to estimate the timescale of the star-formation quenching, as suggested by \citetalias{Citro2017}. 
We define t$_\text{Q}$ as the time elapsed from when the candidate was a typical star forming galaxy to the moment in which it is observed. For QGs-A, this happens when our tracer of the ionisation parameter (i.e. [\ion{O}{III}]/H$\alpha$) becomes $3\sigma$ lower (i.e. about $0.4$ dex) than the median of the [\ion{O}{III}]/H$\alpha$  distribution of SF galaxies. For QGs-B this occurs instead when [\ion{O}{III}]/[\ion{N}{II}]  becomes $3\sigma$ lower than the [\ion{O}{III}]/[\ion{N}{II}] expected from its estimated metallicities.

Firstly, we derive the fraction of the QGs-A and QGs-B as the number of QGs over the number (i.e. 174000) of star-forming galaxies (SF-Alldet plus SF-[\ion{O}{iii}]undet). We obtain a fraction of 0.11\% and 0.18\%, for the QGs-A and QG-B, respectively. 
To obtain the observed quenching timescale of our QGs, following \citetalias{Citro2017}, we multiply this fraction ($F_{QGs}$) by the typical lifetime of a star-forming galaxy, that could be represented by the doubling mass time t$_\text{doubling}$ \citep[i.e. the time needed to a galaxy for doubling its stellar mass ($\sim 1/\text{sSFR}$; e.g.)][]{Guzman1997,Madau2014}: % we obtained the observed quenching timescale of our QGs:
\begin{equation}
\text{t}_\text{Q}= \text{F$_{QGs}$} \times \text{t}_\text{doubling} = \text{F$_{QGs}$} \times \frac{1}{\text{sSFR}}
\end{equation} 
Following the empirical relations by \cite{Karim2011}, we derive the quantity 1/sSFR, which amounts to $\sim 8.8$ and $\sim 10$ Gyr for QGs-A and -B, respectively (assuming a median mass log(M/M$_\odot)\simeq 10.1$ and $10.4$ for -A and -B, respectively).
Then we obtain t$_\text{Q} \sim 9.7 - 18$ Myr.
This t$_\text{Q}$ is a lower limit because of the several conservative assumptions taken into account for the selection of the candidates and because the flux limit of the survey allows to select only the most extreme candidates.

We derive also an upper limit to t$_\text{Q}$ by assuming that about $50\%$ of SF-[\ion{O}{III}]undet population (i.e. $\sim 13000$ galaxies) is in a low ionisation state. 
This is supported by the fact that the [\ion{O}{iii}]/H$\alpha$  ratio in their median stacked spectrum (see \autoref{fig:methodA_summ}) is $1\sigma$ below the median value of SF galaxies (therefore at least $50\%$ of SF-[\ion{O}{III}]undet  are above $1\sigma$ value of SF galaxies, i.e. are consistent to be star-forming galaxies).
In this case, the observed fraction of SF-[\ion{O}{III}]undet is $\sim 7.5\%$ and, with a median log(M/M$_\odot$) $\sim 10.6$ \citep[i.e. 1/sSFR $\sim 10.1$ Gyr, following][]{Karim2011}, the t$_\text{Q}$ is $\sim 0.76$ Gyr, compatible with galaxies which are experiencing a smoother and slower quenching.

Finally, we perform a survival analysis (ASURV, i.e. Kaplan-Meier estimator) of the distribution of [\ion{O}{iii}]/H$\alpha$ in slices of [\ion{N}{ii}]/[\ion{O}{ii}].
We find that $938$ (i.e.  a fraction of  $0.58\%$) among [\ion{O}{iii}]undet galaxies are re-distributed below $3\sigma$ (i.e. the thresholds for QGs-A), representing therefore the global fraction of QGs-A and leading to a quenching timescale of \^t$_\text{Q} \sim 50$ Myr. This time \^t$_\text{Q}$ should represents a good statistical measurement of the true quenching timescale for the adopted threshold. 
With the same ASURV analysis we confirm the consistency of the assumption that about 50\% of [\ion{O}{iii}]undet galaxies are re-distributed below $1\sigma$, accordingly to the value obtained from the median stacked spectra.

We, therefore, convert these values of t$_\text{Q}$s in an \emph{e}-folding time $\tau_\text{Q}$ for the star formation quenching history. 
Adopting the \citetalias{Citro2017} models, we derive the relation between the time needed by the [\ion{O}{III}/H$\alpha$] ratio to decrease by $0.42$ dex and $\tau_\text{Q}$. We find that our lower limit timescales are compatible with an exponential $\tau_\text{Q} \simeq 18-34$ Myr for -A and -B respectively.
Instead, from a linear extrapolation at t$_\text{Q} \sim 0.76$ Gyr we obtain a $\tau_\text{Q} \sim 1.5$ Gyr for the upper limit timescale.
Finally, from \^t$_\text{Q}$ we obtain an estimate of $\hat{\tau}_\text{Q} \sim 90$ Myr.

In summary, from the fraction of our QGs candidates, we derive a broad range of quenching timescales of $10$ Myr $< t_Q <  0.76$ Gyr, and a statistically estimate  of \^t$_\text{Q} \sim 50$ Myr for QGs-A. These values correspond to a range for the \emph{e}-folding time scale of the star formation quenching history of 
$18$ Myr $< \tau_\text{Q} < 1.5$ Gyr
%that corresponds to a 
and an estimate of
$\hat{\tau}_\text{Q} \sim 90$ Myr.

\subsection{Quenching mechanisms}
Our sample of QG candidates is fundamental to get insights on the physical mechanisms driving the quenching of their SF.
From our sample, we find a relatively rapid timescale for quenching (from few Myr to at most 1.5 Gyr) acting in galaxies with log(M/M$_\odot$) > 9.5, being preferentially satellites in intermediate to high-density environments, and having their morphology almost unaffected. 
A smaller fraction ($\sim 25\%$) of our QGs is, however, also in low-density environments, and likely isolated. Only a small fraction ($12-17$\%) of them  have already a compact morphology consistent with a morphological transformation. 
Therefore, different mechanisms should have driven their quenching, in particular in isolated and in high-D environments, and in different evolutionary epochs.

In general, the SFR in the inner star-forming regions of main-sequence galaxies is thought to be fueled through a continuous replenishment of low-metallicity and relatively low-angular momentum gas from the surrounding hot corona \citep{Pezzulli2016}, regulated via stellar feedback \citep[e.g. galactic fountain accretion][]{Shapiro1976,Fraternali2006} until a quenching mechanism shall act to break the process. 
Recently, \cite{Armillotta2016} found that the efficiency of fountain-driven condensation is strictly dependent on the coronae temperature.
In coronae with temperatures higher than $4\times10^6$ K the process is highly inefficient.
Hence, isolated QG candidates which are more massive than the Milky Way could have lost the ability to cool coronae gas and, after the consumption of their gas reservoir, they could start to quench the star-formation.
Instead, in isolated QGs less massive than the Milky Way, strong stellar feedback (i.e. SN and strong stellar wind) may have inhibited the accretion of cold gas from the corona \citep{Veilleux2005,Sokolowska2016} .  

In denser environments there are several quenching mechanisms which can affect our satellite QGs.
However, the most likely process should leave the morphology almost unaffected. 
In dense environment, in which the velocity of galaxies may be sufficiently high, the ram pressure could remove cold gas from the reservoir of the satellite galaxies \citep[e.g.][]{Gunn1972,Quilis2000,Poggianti2004}.
This 'ram pressure stripping' results in a quenching of the star-formation, with a relatively short timescale \citep[$\sim 200$ Myr - $\geq 1$ Gyr, e.g.][]{Steinhauser2016}.
However, there is general consensus that 'strangulation' (or 'starvation') is the dominant quenching mechanism in satellite galaxies \citep[e.g.][]{Treu2003,VanDenBosh2008,Peng2015}.
When the corona of a galaxy interacts hydro-dynamically with the hot and dense intra-cluster medium of a larger halo, its hot and diffuse gas could be stripped \citep[e.g.][]{Larson1980,Balogh2000}.
This effect leads to a suppression of the accretion onto the disk, thus resulting in a gradual decline of the SFR until the exhaustion of the gas reservoir of the galaxy.  
\cite{Peng2015} claimed that the primary quenching mechanism for galaxies is the strangulation. They derive the quenching timescale due to this mechanism as the gas depletion time needed to explain the differences in metallicity between the star-forming population and the quiescent one, finding that models of $4$ Gyr are the most suitable for this task \citep[see also][]{Maier2016}.
We compare this timescale with the estimate of the time (t$_\text{P}$) needed by a galaxy to decrease its sSFR from typical main sequence values \citep[i.e. $\sim -10$ in log-scale for log(M/M$_\odot)\sim 10.4$, e.g.][]{Karim2011} to the typical sSFR of the passive population.
 \citep[i.e. $\sim -11$ in log scale, e.g.][]{Pozzetti2010, Ilbert2013}.
For the \emph{e}-folding time $\tau_\text{Q}$ we derive a lower limit timescale t$_\text{P}$ of $\sim 40-80$ Myr and an upper limit of $\sim 3.5$ Gyr. 

Recently, it is gaining consensus a scenario \citep[i.e. 'Delayed then rapid' quenching][]{Wetzel2013,Fossati2017} in which the quenching of satellites in dense environments has been proposed to be divided into two phases: a relatively long period (``the delay time'') in which the star-formation does not differ strongly from the main-sequence values (2-4 Gyr after first infall), followed by a phase in which the SFR drops rapidly (``the fading time'') with an exponential fading with an e-\emph{folding} time of $0.2 - 0.8$ Gyr (lower for more massive galaxies) that is independent on host halo mass. 

About $70\%$ and $80\%$ of QGs-A and -B reside in high- and intermediate- density environments and the vast majority of them are satellites. Hence, the derived $\tau_\text{Q}$ for our QGs allow to perform a direct comparison with the quenching timescales in \cite{Wetzel2013}.
Although the range derived for $\tau_\text{Q}$ is quite broad, we can at least confirm that our timescales are compatible with theirs. 

We can also make a further comparison with the \cite{Wetzel2013} predictions by using the median L(H$\alpha$) of the stacked spectra (see \autoref{sec:spec_prop}) as a proxy of the median SFR, and comparing them to the value of the SF galaxy sample (0.8 for QGs-A and 0.9 for QGs-B).
Assuming that, after the start of the quenching, the SFR of a star-forming galaxy decreases to the value of SFR$_\text{QGs}$(t$_\text{Q}$), we can estimate the exponential \emph{e}-folding time, deriving $\tau_\text{Q} \sim 50 - 190$ Myr for the median QGs-A and -B spectra, that are closer to those found by \cite{Wetzel2013}.

It is also interesting to limit the discussion to the central quenching galaxies.
There is consensus that the quenching is slower for central galaxies \citep[e.g.][]{Hahn2017}.
Focusing on central galaxies in high density environment we found a lower limit t$_\text{Q}$ of $8.8-20$ Myr respectively for central QGs-A and -B and an upper limit t$_\text{Q} \sim 1.4$ Gyr. These timescales are compatible with lower limit exponential e-folding $\tau_\text{Q}$	of $16-38$ Myr and an upper limit $\tau_\text{Q} \sim 2.8$ Gyr. Following an approach similar to that of \cite{Wetzel2013}, \cite{Hahn2017} found that central galaxies quench the star formation with an \emph{e}-folding time between $0.5$ and $1.5$ Gyr (lower for massive galaxies) and also in this case we are compatible with their result.

There is general consensus that environmental mechanisms take longer time ($2-3$ Gyr, \citep[e.g.][]{Balogh2000, Wang2007} to start to affect the SFR of satellites. Hence, this should be the most probably scenario also for our satellites before the starting of the quenching.
Moreover, we found that our quenching timescales are compatible with an exponential decrement like that of \cite{Wetzel2013}, although without an overwhelming.
We can conclude that the properties of our QG candidates can be preferentially explained by a 'Delayed then rapid' quenching  of satellites galaxies, due to the final phase of an environmental quenching mechanism(s).

Strangulation, ram-pressure stripping and harassment \citep[e.g.][]{Farouki1981} are the most important mechanisms that act on satellites leading to the quenching of their star formation.
Since we can safely exclude from a visual inspection that any of our candidates are experimenting harassment, the natural conclusion for our QGs that reside in high- and intermediate- density environment is that strangulation and ram-pressure stripping should play a primary role in the halt of the star-formation. 
However, different mechanisms can also work together to quench the star-formation in galaxies. 
Moreover, even if our methods are not sensitive to AGN feedback \citep[e.g.][]{DeLucia2006,Fabian2012,Cimatti2013,Cicone2014} as quenching mechanism, due to our a priori exclusion of AGNs, we cannot exclude that an early AGN phase could be responsable and could have quenched our QGs. In this case, the AGN phase should have finished before the quenching of the star formation.
On the other end, it is important to remind that other authors predict the formation of ETGs without invoking the AGN feedback \citep[e.g.][]{Naab2006,Johansson2012}.

Finally, we stress that our QGs still have a ionised-gas phase, as witnessed by their H$\alpha$ emission, even if weaker than in the parent sample of star forming galaxies, suggesting that gas depletion is indeed on going.
In the future, after the disappearance of late-B stars and the consequently disappearance of hard-UV photons, this gas could be cooled down being available for a new phase of star formation.
However, if we are witnessing a minimum in the SFH of our quenching candidates, we should observe QGs at all masses, also lower than $10^{9.5} \text{M}_\odot$ (i.e. the less massive QG in our sample). 
To confirm the final passive fate of our QGs, connected to the presence or absence of residual gas and, at the same times explaining the observed H$\alpha$ emission, more observations are needed and in particular we need to study the cold gas phase distribution from ALMA observations \citep[see, for instance,][]{Decarli2016,Lin2017}.

%%%%%%%%%%%%%%%%%%%%%%%%%%%%%%%%%%%%%%%%%%%%%%%%%%
%%%%%%%%%%%%%%%%%%%%%%%%%%%%%%%%%%%%%%%%%%%%%%%%%%
%%%%%%%%%%%%%%%%%%%%%%%%%%%%%%%%%%%%%%%%%%%%%%%%%%

%%%%%%%%%%%%%%%%%%%%%%%%%%%%
% Conclusion
%%%%%%%%%%%%%%%%%%%%%%%%%%%%

\section{SUMMARY}
In this work, we analyse a sample of $\sim$ 174000 SDSS-DR8 star-forming galaxies at $0.04 \leq \text{z} < 0.21$ to provide for the first time a sample of quenching galaxy (QGs) candidates selected just after the interruption of their star formation.

We follow the approach introduced by \citetalias{Citro2017} to select QG candidates on the basis of emission line flux ratios of higher-to-lower ionisation lines (i.e. [\ion{O}{iii}]/H$\alpha$) which are sensitive to the ionisation level.  
The main issue of this approach is that the [\ion{O}{iii}]/H$\alpha$ ratio is affected by a ionisation-metallicity degeneracy. 

In order to mitigate this degeneracy we set up two different methods:

\begin{itemize}
\item \emph{Method A}: following \citetalias{Citro2017}, we exploit %the [\ion{N}{II}]/[\ion{O}{ii}] ratio as metallicity indicator, in 
the plane [\ion{O}{iii}]/H$\alpha$ vs. [\ion{N}{II}]/[\ion{O}{ii}] to select galaxies with the lowest [\ion{O}{iii}]/H$\alpha$ values for a given metallicity (i.e. using [\ion{N}{II}]/[\ion{O}{ii}] as metallicity indicator). %fixed metallicity). 
By analysing the statistical distribution of our samples, we identify an excess of galaxies consistent with being a population separated from the star-forming sample, having intrinsically lower [\ion{O}{iii}]/H$\alpha$ values, and hence lower ionisation levels.
This method is demonstrated to be stable against the choice of the dust attenuation law.
We also tested an alternative diagram involving 
[\ion{O}{iii}]/H$\beta$ vs. [\ion{N}{II}]/[\ion{S}{ii}], which has the advantage of being less affected by dust extinction, although it involves weaker lines. 

\item \emph{Method B}: an alternative method to mitigate the metallicity degeneracy is to select galaxies for which the [\ion{O}{iii}] flux is weaker than the minimum value expected from their metallicity, exploiting the two relations [\ion{N}{II}]/[\ion{O}{ii}] vs. 12+log(O/H) and 12+log(O/H) vs. [\ion{O}{iii}]/[\ion{N}{II}].
\end{itemize}

Applying these methods we select two samples of QG candidates and analyze their main properties. Our results can be summarized as follows:

\begin {enumerate}
\item We select 192 candidates (QGs-A) and 308 candidates (QGs-B), using Method A and B, respectively. There is an intersection of 120 QGs between them, and the QGs out of the intersection are close to the threshold criterion of both methods. QGs-B show, on average, higher values of [\ion{N}{II}]/H$\alpha$ and of [\ion{N}{II}]/[\ion{O}{II}] compared to QGs-A, suggesting they have a statistically higher metallicity than QGs-A.

\item The median stacked spectra, corrected for dust extinction, of QGs-A and QGs-B have a blue stellar continuum, suggesting a young mean stellar population.
The analysis confirms a weak [\ion{O}{iii}] emission and that also other high ionisation emission lines (such  as [\ion{Ne}{iii}]) are weak in both QGs stacked spectra, and the [\ion{O}{iii}]/H$\alpha$ ratios are consistent with a low ionisation level. 
On the contrary, [\ion{O}{ii}], H$\alpha$ and H$\beta$ (i.e. low ionisation lines) are still strong and  with an H$\alpha$ flux slightly weaker ($\sim 80 \%$) than the one of star-forming galaxies.

\item We find that QGs have D$_\text{n}$4000 and EW(H$\alpha$)  values intermediate between SF and already quenched galaxies, confirming that they have just stopped their star formation and have a young/intermediate mean stellar population.

\item In the colour-mass diagram, the bulk of the QGs-A and QGs-B resides in the blue cloud region, and only few of them ($\sim 3$\%) lie in the green valley region.
The QGs have masses log(M/M$_\odot$) > 9.5, comparable with those of the SF population, being the QGs-B, on average, more massive. 
This suggests that, as expected in a downsizing scenario, star formation quenching has not started yet for low-mass galaxies, consistently with a lack of low-mass red galaxies.
Their H$\alpha$ emission, is compatible with a just quenched 
SFR of the order of 0.6 - 10 M$_\odot$ yr$^{-1}$,  similar to that of the SF main sequence population, suggesting the presence of residual ionised gas in our QGs. The emission from the median stacked spectra is, however, weaker than in the SF population, suggesting that the depletion of the gas has started.

\item The morphology and concentration index (C=R90/R50) of QGs are similar to those of the star-forming population, suggesting that no morphological transformation has occurred yet, in the early phase after the quenching of the star formation. 
However, some of them have a concentration index higher than the threshold that divides the early-type from the other galaxy types (i.e. C > 2.6) and they could have experienced morphological transformation during the interruption of the star formation.

\item Compared to the parent SF population, we find an excess of QGs in high density environments ($\sim$ 42\%), in particular for QGs-B. %\sout{a lack of QGs in low-D ($\sim$ 25\%)}. 
QGs in high density environments are preferentially satellites (from $\sim $ 60 to 80\%). %, while only a low fraction ($\sim$ 4 to 7 \%) are satellites in low-D environments. 
Approximatively 5\% of QGs are in groups/clusters which have more than 30 members, but no one of them is the brightest galaxy in its environments.

\item From the fraction of QG candidates  ($\sim 0.11 - 0.18\%$ of the SF population) we estimate the quenching timescales for these populations to be between 10 and 18 Myr.
These values are compatible with the sharp quenching models by \citetalias{Citro2017}.
If we assume that at most 50\% the entire SF-[\ion{O}{iii}]undet population ($\sim$ 7.5\%) is in a low ionisation state, as witnessed by the low but not extreme [\ion{O}{iii}]/H$\alpha$ ratio in their stacked spectrum, we obtain an upper limit to the quenching timescale of $\simeq$ 0.76 Gyr.
In this case, the quenching timescale is compatible with galaxies which are experiencing a more smoothed and slower quenching (\citetalias{Citro2017}).  We convert this range into a e-folding timescale for the SFR quenching history, finding 18 Myr $< \tau_\text{Q} < 1.5$ Gyr.

\item  From a survival analysis we find that 938 ($\sim 0.58\%$) among SF-[\ion{O}{iii}]undet galaxies are consistent to be QGs candidates. We, therefore, derive a statistical measurement of the quenching timescale \^{t}$_\text{Q}$ of $\sim 50$ Myr and $\hat{\tau}_\text{Q} \sim 90$ Myr.

\end{enumerate}

This analysis, based on a new spectroscopic approach, leads to the  identification of a population of galaxy candidates selected right after the quenching.
Most of our QGs would not have been selected as an intermediate population using colour criteria, or in the SFR-mass plane.

We conclude that these galaxies, that  are quenching their star formation on a short timescale (from few Myr to less than 1 Gyr), preferentially reside in intermediate-to high-density environments, are satellites and have not morphologically transformed into spheroidal red passive galaxies yet. All these properties  could be explained by a 'Delayed then rapid' \citep[see][]{Wetzel2013} quenching scenario in satellites galaxies, due to the final phase of strangulation or ram pressure stripping.

However, to confirm the proposed scenario and the presence or absence of a reservoir of gas more observations are needed. For example, cold gas phase distribution could be derived from ALMA observations \citep[e.g.][]{Decarli2016,Lin2017}, while the spatial distribution of quenching can be analyzed using integral field unit (IFU) spectroscopic data to get insights about the inside-out scenario (\citealp[e.g. the MANGA public survey,][]{Bundy2015}; \citealp[the SAMI galaxy survey,][]{Green2018}; \citealp[the MUSE-VLT data][]{Bacon2010} \citealp[and, in the future, WEAVE-IFU data,][]{Dalton2014}).

%\cleardoublepage
\section*{Acknowledgements}
The authors thank the anonymous referee for helpful suggestions and very constructive comments. We are grateful to Filippo Fraternali, Paola Popesso, Gianni Zamorani, Roberto Maiolino, Christian Maier, Filippo Mannucci, Sirio Belli and Alice Concas for useful discussion and suggestions and Cristian Vignali to help us for the ASURV analysis.
The authors also acknowledge the grants ASI n.I/023/12/0 ``Attivit\`a relative alla fase B2/C per la missione Euclid'' and PRIN MIUR 
2015 ``Cosmology and Fundamental Physics: illuminating the Dark Universe with Euclid''.
Funding for the SDSS and SDSS-II has been provided by the Alfred P. Sloan Foun- dation, the Participating Institutions, the National Science Foundation, the U.S. Department of Energy, the National Aeronautics and Space Administration, the Japanese Mon- bukagakusho, the Max Planck Society, and the Higher Ed- ucation Funding Council for England. The SDSS Web Site is http://www.sdss.org/. The SDSS is managed by the As- trophysical Research Consortium for the Participating In- stitutions. The Participating Institutions are the American Museum of Natural History, Astrophysical Institute Pots- dam, University of Basel, University of Cambridge, Case Western Reserve University, University of Chicago, Drexel University, Fermilab, the Institute for Advanced Study, the Japan Participation Group, Johns Hopkins University, the Joint Institute for Nuclear Astrophysics, the Kavli Insti- tute for Particle Astrophysics and Cosmology, the Korean Scientist Group, the Chinese Academy of Sciences (LAM- OST), Los Alamos National Laboratory, the Max-Planck- Institute for Astronomy (MPIA), the Max-Planck-Institute for Astrophysics (MPA), New Mexico State University, Ohio State University, University of Pittsburgh, University of Portsmouth, Princeton University, the United States Naval Observatory, and the University of Washington.

%%%%%%%%%%%%%%%%%%%%%%%%%%%%%%%%%%%%%%%%%%%%%%%%%%



\bibliographystyle{mnras}
\bibliography{biblio_Quench} 

\appendix
\section{Fibre aperture effects}
\begin{figure}
   \centering
 \includegraphics[width=\columnwidth]{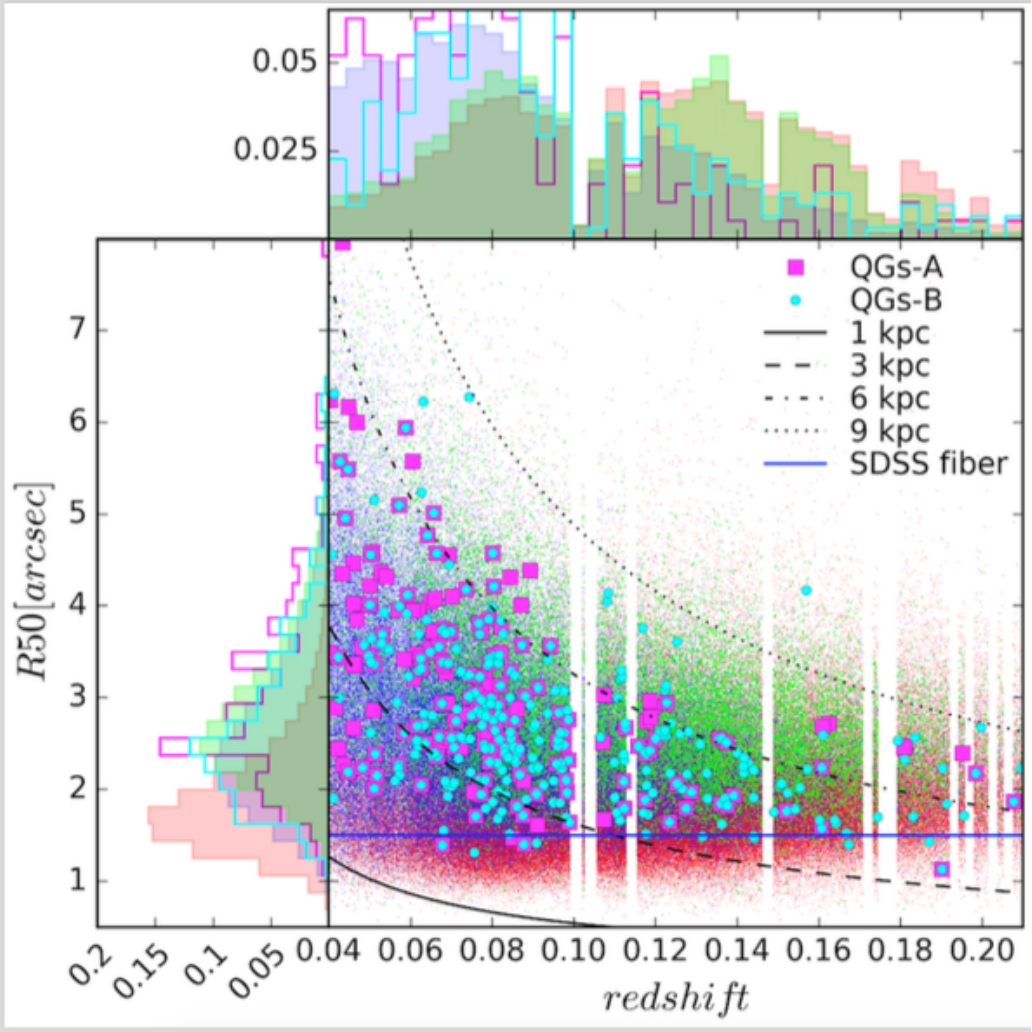} \hfill
   \caption{The size-redshift relation for our sample and quenching candidates. The horizontal blue line represents the radius of the SDSS fibre aperture, while the black curves represent the kpc/arcsec relations obtained from the adopted cosmology.}
         \label{fig:r50_z}
\end{figure}
\begin{figure}
   \centering
   \includegraphics[width=\columnwidth]{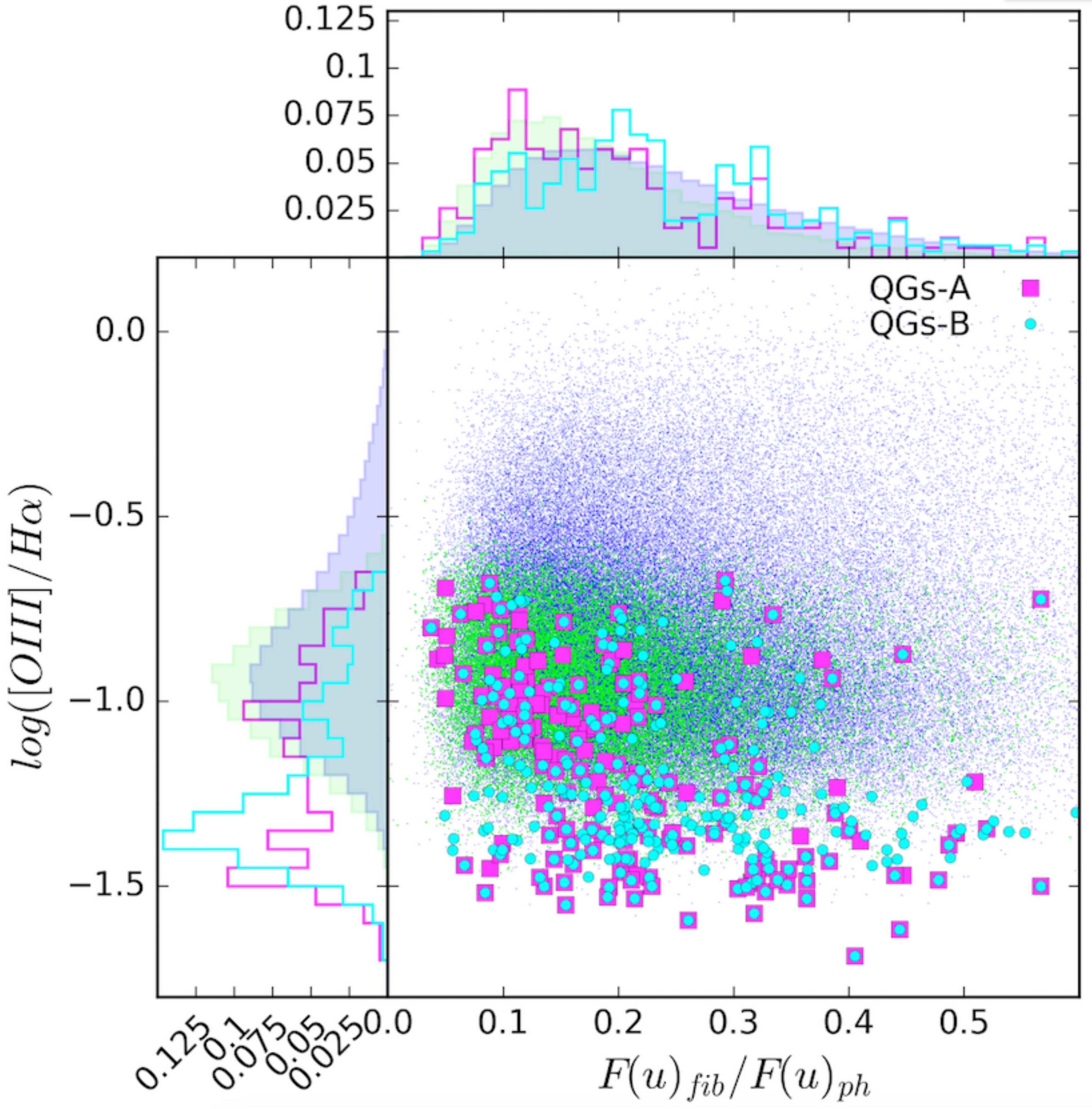} \hfill
   \caption{[\ion{O}{iii}]/H$\alpha$ as a function of the ratio between the u-band flux inside the fibre and total. The colour code is the same of \autoref{fig:methodB_summ}.}
         \label{fig:O3Ha_apert}
\end{figure}
As mentioned in \autoref{subsec:sampsel}, we select galaxies at z $> 0.04$, following the prescription of \cite{Kewley2005}, in order to avoid strong fibre aperture effects on SFRs, extinction and metallicity. 
Indeed, we stress that our analysis is limited %\sout{to provide constraints on the mean quenching status of our galaxies} 
to the area of galaxies covered by the fibre.
With the adopted cosmology, at z $= 0.04$ and z $=0.21$, the SDSS fibre radius (1.5 arcsec) corresponds to $\sim 1.2$ kpc and $\sim 5.2$ kpc, respectively.
In \autoref{fig:r50_z} we show the size-redshift relation for our sample. 
The size is represented by the R50 radius. 
Roughly all the candidates have R50 larger than the fibre radius. 
Therefore, we can analyse the quenching only in the inner part of our QG candidates. 
In order to test whether we could extend our results to the whole galaxy,  we explore the impact of the aperture on the ionisation indicator [\ion{O}{iii}]/H$\alpha$. 
\autoref{fig:O3Ha_apert} shows the [\ion{O}{iii}]/H$\alpha$ as a function of the fraction of  flux inside the fibre (u-band) with respect to the total u-band flux, \citep[u-band Petrosian flux,][]{Strauss2002}.
There is no evident trend, with QGs candidates distributed over the whole range,  
%gradient of ionisation indicator, 
even at the lowest ionisation levels (log([\ion{O}{iii}]/H$\alpha) < -1.3$). 
This test suggests that, in this redshift range, the fibre aperture does not affect significantly our analysis, even if our results are clearly relative only to the region included in the fibre. %{\color {red} Finally, \autoref{fig:r50_z} reveals that there is a discrepancy in the redshift of QGs and SF-Alldet and SF-[\ion{O}{ III}]undet. Even if the median redshifts agree ($0.077 \pm 0.003$ for QGs-A, $0.084 \pm 0.03$ for QGs-B and $0.0801 \pm 0.0001$ for SF-Alldet), the redshift distributions of the Since at low redshift the volume limit leads to observe less massive galaxies, these discrepancies should be explained by the difference in their mass distribution (see e.g. \autoref{fig:160117_3} and \autoref{tab:060217_1}).}

%\begin{thebibliography}{99}
%\end{thebibliography}

%%%%%%%%%%%%%%%%%%%%%%%%%%%%%%%%%%%%%%%%%%%%%%%%%%

%%%%%%%%%%%%%%%%% APPENDICES %%%%%%%%%%%%%%%%%%%%%

%\appendix



%%%%%%%%%%%%%%%%%%%%%%%%%%%%%%%%%%%%%%%%%%%%%%%%%%

% Don't change these lines
\bsp	% typesetting comment
\label{lastpage}
\end{document}